\documentclass[aps,prl,twocolumn,showpacs,superscriptaddress,groupedaddress]{revtex4}  
\usepackage{graphicx}  
\usepackage{dcolumn}   
\usepackage{bm}        
\usepackage{amssymb}   
\usepackage{color}

\newcommand{\COMMENTED}[1]{}

\begin{document}

\widetext

\title{Finite-Temperature Auxiliary-Field Quantum Monte Carlo for Bose-Fermi Mixtures}
\author{Brenda M. Rubenstein}
\affiliation{Department of Chemistry, Columbia University, New York, NY 10027}
\author{Shiwei Zhang}
\affiliation{Department of Physics, College of William and Mary, Williamsburg, VA 23187-8795}
\author{David R. Reichman}
\affiliation{Department of Chemistry, Columbia University, New York, NY 10027}
\date{\today}

\begin{abstract}
We present a quantum Monte Carlo (QMC) technique for calculating the exact finite-temperature properties of Bose-Fermi mixtures. The Bose-Fermi Auxiliary-Field Quantum Monte Carlo (BF-AFQMC) algorithm combines two methods, a finite-temperature AFQMC algorithm for bosons and a variant of the standard AFQMC algorithm for fermions, into one algorithm for mixtures. We demonstrate the accuracy of our method by comparing its results for the Bose-Hubbard and Bose-Fermi-Hubbard models against those produced using exact diagonalization for small systems. Comparisons are also made with mean-field theory and the worm algorithm for larger systems. As is the case with most fermion Hamiltonians, a sign or phase problem is present in BF-AFQMC. We discuss the nature of these problems in this framework and describe how they can be controlled with well-studied approximations to expand BF-AFQMC's reach.  The new algorithm can serve as an essential tool for answering many unresolved questions about many-body physics in mixed Bose-Fermi systems. 

\end{abstract}

\maketitle


\section{\label{sec:introduction}I. Introduction}

Ultracold atomic gases loaded into optical traps offer the unique possibility of experimentally simulating many of the fundamental models of condensed matter physics \cite{ColdAtomBloch, ColdAtomLewenstein}. These systems are clean, and owing to remarkable advances in trapping, cooling, and the manipulation of inter- and intraparticle interactions,  may be studied with an unprecedented level of experimental control. One of the field's landmark achievements has been the observation of the superfluid-Mott insulator transition in Bose gases \cite{SFMI}. Analogous successes with fermions have led to the direct observation of such phenomena as Fermi pressure and antibunching \cite{ColdAtomFermionsTruscott, ColdAtomFermionsRom}. Focus has now shifted to ultracold mixtures of bosons and fermions \cite{Inouye, Goldwin, Ospelkaus1, Ospelkaus2, Ospelkaus3, Ospelkaus4, Gunter}. At the most practical level, bosons may be used to sympathetically cool trapped fermions \cite{Hadzibabic, Schreck}. Much more tantalizing, however, is the prospect that bosons may be able to mediate a BCS superfluid transition in ultracold Fermi gases \cite{Heiselberg, Efremov, Illuminati}, or emulate many-body Hamiltonians of mixture systems predicted to exhibit a plethora of exotic phases \cite{BFPhaseDiagramLewenstein, BFPhaseDiagramFehrmann}. Equally intriguing is the possibility of using newly created ``Bose-Fermi molecules'' with permanent dipole moments as qubits for quantum computers or as probes of the permanent electric dipole moment of the electron \cite{Zirbel, Ospelkaus3, DeMille, OspelkausReview}. These possibilities have galvanized both experimentalists and theorists to develop new tools capable of exploring the full range of mixture phenomenology. 

From a theoretical standpoint, delineating the exact finite-temperature Bose-Fermi phase diagram represents a formidable challenge. Mean-field and perturbation theory calculations suggest that Bose-Fermi mixtures may exhibit a wide variety of behaviors, ranging from Bose-Fermi ``molecule'' spin and charge density waves to phase segregation \cite{BFPhaseDiagramLewenstein, BFPhaseDiagramFehrmann, Buchler, Kuklov, Albus, Cramer, Mathey}. Nevertheless, these techniques are approximate by definition, which raises concerns about the phase diagrams they yield. A reliable description of Bose-Fermi mixture phenomenology requires an exact framework capable of accurately accounting for strong correlation among particles. Accurate results can be obtained for small clusters whose limited Hilbert spaces are amenable to exact diagonalization (ED), and linear chains for which quantum Monte Carlo (QMC) techniques free of the sign problem or density matrix renormalization group methods may be applied \cite{Zujev, Pollet, Varney, Hebert1, Hebert2, Schollwock}. Techniques for large systems in two and higher dimensions, however, are scarce. 

The most promising and flexible technique for mixtures to date uses the framework of Dynamical Mean Field Theory (DMFT) \cite{Kotliar}. While initial applications of DMFT to mixtures paired well-established DMFT methods for fermions with approximate treatments of bosons \cite{Byczuk, Titvinidze1, Titvinidze2}, the first rigorous Bose-Fermi DMFT algorithm has recently been proposed, which weds fermion DMFT with a newly-derived DMFT approach for bosons \cite{AndersBFDMFT, AndersBDMFT1, AndersBDMFT2}. As with all DMFT approaches, this technique is only expected to be accurate in the limit of large dimensionality or coordination number. Indeed, recent Boson-DMFT (BDMFT) calculations on the Bose-Hubbard model demonstrate that, while DMFT is remarkably accurate in three dimensions, it is less so in two dimensions \cite{AndersBDMFT2}. Furthermore, because DMFT is most useful for systems with short-range correlations, inhomogeneous phases and long-wavelength collective modes may present additional challenges. 

In contrast, QMC  techniques offer the promise of being exact regardless of system size, dimensionality, and homogeneity. QMC techniques differ widely in detail from algorithm to algorithm, but all employ stochastic sampling to solve the Schr\"{o}dinger equation at zero temperature or determine  partition and correlation functions at finite temperatures. Because of their accuracy and modest computational cost, QMC methods such as the worldline and worm algorithms have become the techniques of choice for boson lattice models \cite{Batrouni1, Batrouni2, Prokofev1, Prokofev2}. Auxiliary-field and diagrammatic QMC techniques also exist for fermions \cite{BSS1, BSS2, BSS3, Phaseless, Prokofev3, VanHoucke}. Unlike techniques for bosons, however, fermion QMC in two or more dimensions is generally plagued by the sign problem, resulting in an exponential scaling of computational cost with inverse temperature to achieve a fixed accuracy \cite{SignProblem}. Developing a widely-applicable QMC technique for mixtures thus requires not only marrying two considerably different fermion and boson techniques together, but finding a way to tame the sign problem within that combined formalism. 

Widely employed in condensed matter and nuclear physics, the Auxiliary-Field Quantum Monte Carlo (AFQMC) method \cite{BSS1,Sugiyama, Bai}  is a field theoretical method where many-body propagators resulting from two-body interactions are transformed into many-dimensional integrals over one-body propagators using the Hubbard-Stratonovich Transformation \cite{HSTransform1, HSTransform2}. The resulting integrals are then computed using Monte Carlo sampling. In recent years, AFQMC has predominantly been used to study the equilibrium properties of the Hubbard model both at finite-temperature and in the ground-state. Like all fermion QMC techniques, conventional AFQMC suffers from the sign problem in most parameter regimes. However, an alternative formulation, in which walkers are pruned using population control techniques as they sample AFs in imaginary time, has allowed a general, efficient approach to treat both local and extended interactions. This framework allows the constrained-path and phaseless approximations to be easily incorporated to control the sign and phase problems \cite{CPMC1, CPMC2,Phaseless, Chang2008}. In recent years, these approximations have been tested on a variety of systems including the Hubbard model \cite{CPMC1, Phaseless, Chang2008} and the electronic structure of solids and molecules \cite{alsaidi1, alsaidi2} and has been shown to yield accurate energies and correlation functions. Thus, Constrained-Path AFQMC (CPMC) is well-equipped to explore phases beyond the scope of other fermion QMC methods. The formalism of AFQMC has also previously been generalized to treat bosons in the ground state \cite{Purwanto1, Purwanto2}. This suggests that AFQMC would be perfectly suited for studying mixtures  via a combination of bosonic and fermionic Monte Carlo techniques if the formalism could be further expanded to treat bosons at finite temperatures. 

In this work, we present an exact QMC methodology that can be used to determine the thermodynamic properties of Bose-Fermi mixtures in any dimension over a wide range of parameters.  Our method, Bose-Fermi Auxiliary Field Quantum Monte Carlo (BF-AFQMC), generalizes finite-temperature AFQMC for fermions to bosons and Bose-Fermi mixtures. By casting the bosonic portion of the problem in terms of auxiliary fields, we can extend determinantal QMC techniques to bosons and sample the boson partition function by sampling determinants just as one would for fermions. We arrive at an exact technique for mixtures by combining our approach for bosons with previous AFQMC techniques for fermions. We then discuss how the constrained path and phaseless approximations can be imposed to remove the sign and phase problems in our method. As a benchmark, we compare our algorithm's results for Bose-Hubbard and spin-polarized Bose-Fermi-Hubbard clusters to those obtained using ED. We also contrast our results with those from mean-field theory (MFT) and the worm algorithm. 

Our paper is organized as follows: In Section II, we begin by reviewing the AFQMC formalism for fermions as background for our new algorithm. We then proceed to present the underlying formalism for our new boson and Bose-Fermi algorithms in Section III, including importance sampling schemes. We also outline the implementation of the constrained path and phaseless approximations, which can respectively control the sign and phase problems. In Section IV, we compare our algorithm's results for the Bose-Hubbard and spin-polarized Bose-Fermi Hubbard models against those produced using alternative methods in an effort to demonstrate the accuracy of our technique.  We finally conclude in Section V, leaving the derivation of the expression relating the boson partition function to a determinant and other details to the appendices. 

\section{\label{Preliminaries}II. Preliminaries}

\section{\label{sec:level2} A. Generic Mixture Hamiltonian and Definitions}

To facilitate the subsequent discussion, we use the following form of the Bose-Fermi-Hubbard Hamiltonian as a concrete example 
\begin{equation}
\hat{H}_{bf} = \hat{K}_{b} +  \hat{K}_{f} + \hat{V}_{b} + \hat{V}_{f} + \hat{V}_{c} , 
\label{SimplifiedBFHEquation}
\end{equation}
where $\hat{K}_{b}$ contains all one-body boson terms
\begin{eqnarray}
\hat{K}_{b} &=& -t_b \sum_{ \langle ij \rangle} \left( \hat{b}_i^{\dagger} \hat{b}_j + H.c. \right) + \sum_{i} \epsilon_{i}^{b}  \hat{n}_{i} , 
\label{eq:Kb}
\end{eqnarray}
$\hat{K}_{f}$ contains all one-body fermion terms
\begin{eqnarray}
\hat{K}_{f} &=&   -t_f \sum_{ \langle ij \rangle, \sigma } \left( \hat{f}_{i\sigma}^{\dagger} \hat{f}_{j\sigma} + H.c. \right) +  \sum_{i, \sigma} \epsilon_{i,\sigma}^{f} \hat{m}_{i,\sigma},
\label{eq:Kf}
\end{eqnarray}
$\hat{V}_{b}$ contains two-body boson terms, 
\begin{eqnarray}
\hat{V}_{b} &=& \frac{U_{b}}{2} \sum_{i} \hat{n}_{i}^{2}, 
\label{eq:Vb}
\end{eqnarray}
$\hat{V}_{f}$ contains two-body fermion terms
\begin{eqnarray}
\hat{V}_{f} &=& U_{f} \sum_{i} \hat{m}_{i\uparrow} \hat{m}_{i\downarrow},  
\label{eq:Vf}
\end{eqnarray}
and $\hat{V}_{c}$ represents the Bose-Fermi coupling term
\begin{equation}
\hat{V}_{c} = C \sum_{i} \hat{n}_{i} \hat{m}_{i}. 
\label{CouplingEquation}
\end{equation}
In the above, $\hat{b}_{i}^{\dagger}, \hat{b}_{i}$ denote the boson creation and annihilation operators and $\hat{f}_{i\sigma}^{\dagger}, \hat{f}_{i\sigma}$ the fermion creation and annihilation operators with spin $\sigma$ ($=\uparrow$ or $\downarrow$) at site $i$. We define the boson density at site $i$ as $\hat{n}_{i} \equiv \hat{b}_{i}^{\dagger} \hat{b}_{i}$ and the fermion densities as $\hat{m}_{i\sigma} \equiv \hat{f}_{i\sigma}^{\dagger} \hat{f}_{i\sigma}$. The total fermion density at each site is denoted by $\hat{m}_{i} \equiv \hat{m}_{i, \uparrow} + \hat{m}_{i, \downarrow}$. $t_{b}$ and $t_{f}$ represent the respective boson and fermion hopping parameters.  $U_{b}$ is the two-body boson-boson potential, $U_{f}$ is the two-body fermion-fermion potential, and $C$ is the Bose-Fermi coupling. $\epsilon_{i}^{b}$ and $\epsilon_{i, \sigma}^{f}$ represent coefficients of one-body terms that may include contributions from chemical potentials, external traps, or disorder. Depending upon the values of the various parameters, this Hamiltonian can exhibit the full range of Bose-Fermi phenomenology. More general Hamiltonians may be handled by the approach outlined below. 

\section{\label{sec:level2} B. Finite-Temperature AFQMC for Fermions}

The finite-temperature AFQMC method for fermions calculates the thermodynamic properties of a system of particles with two-body interactions by reexpressing two-body propagators as integrals over one-body propagators and a set of auxiliary fields. Here, we review the basic formalism to acquaint the reader with previous work relevant to the following discussion \cite{ZhangReview}. In general, the finite-temperature expectation value of an observable, $\hat{O}$, may be written as
\begin{equation}
\langle \hat{O} \rangle \equiv \frac{ Tr\left( \hat{O} e^{-\beta \hat{H}} \right) } {Tr \left( e^{-\beta \hat{H}} \right) }, 
\end{equation}
where $\hat{H}$ is the Hamiltonian of the system and $\beta=1/k_{B}T$. One may rewrite the partition function, $Z$, in terms of a product of $l$ short-time propagators
\begin{equation}
Z = Tr \left( e^{-\beta \hat{H}} \right) = Tr \left( e^{-\Delta \tau \hat{H}} e^{-\Delta \tau \hat{H}}... e^{-\Delta \tau \hat{H}} \right). 
\label{PartitionFunction}
\end{equation}
Here, $\Delta \tau \equiv \beta/l$ is the timeslice in imaginary time. For simplicity, consider the fermion Hamiltonian $\hat{H}_{f} = \hat{K}_{f} + \hat{V}_{f}$ using the definitions from Part A.  One may next perform a Trotter-Suzuki factorization on each of the short-time propagators \cite{SuzukiTrotter1, SuzukiTrotter2}. At second order this yields
\begin{equation}
e^{-\Delta \tau ( \hat{K}_{f} + \hat{V}_{f} ) } = e^{- (1/2) \Delta \tau \hat{K}_{f} } e^{- \Delta \tau \hat{V}_{f} } e^{- (1/2) \Delta \tau \hat{K}_{f}} + O(\Delta \tau^{3}), 
\end{equation}
which becomes exact in the limit $\Delta \tau \rightarrow 0 $. Each short-time propagator is thus a product of two one-body propagators and one two-body propagator. 
In our Hamiltonian, 
\begin{eqnarray} 
\hat{V}_{f} &=& U_{f} \sum_{i} \hat{m}_{i\uparrow} \hat{m}_{i\downarrow} \label{Vf_sqform2} \\
&=&  -\frac{U_{f}}{2} \sum_{i} \left( \hat{m}_{i\uparrow} - \hat{m}_{i\downarrow} \right)^{2} + \frac{ U_{f} }{2} \sum_{i} \left( \hat{m}_{i\uparrow} + \hat{m}_{i\downarrow} \right). \nonumber 
\end{eqnarray}
This form allows the two-body propagators to be reexpressed in terms of an integral over a product of one-body propagators and a set of auxiliary fields using the Hubbard-Stratonovich (HS) Transformation \cite{HSTransform1, HSTransform2}:
\begin{equation}
e^{(1/2) \Delta \tau \hat{v}^{2}} = \frac{1}{\sqrt{2 \pi}} \int_{-\infty}^{\infty} d\phi e^{-(1/2) \phi^{2}} e^{\phi \sqrt{\Delta \tau} \hat{v}},  
\label{HSTransformEquation}
\end{equation}
where $\phi$ is an auxiliary-field (AF). Note that, while there are discrete versions of the HS transformation for the form of $\hat{V}_{f}$ in our Hamiltonian, we have outlined a continuous  version which formally resembles the transformation we will use for $\hat{V}_{b}$ and $\hat{V}_{c}$.

This expression for the short-time propagator may be further simplified by viewing the collection of fields at each timeslice as a vector of fields, $\vec{\phi} \equiv \{ \phi_{1}, \phi_{2}, ..., \phi_{N}\}$, where $N$ is the number of lattice sites, and the normalized Gaussians at each site as probabilities, $p(\phi_{i})$. Collecting all one-body operators into $\hat{B}_f(\vec{\phi})$, we arrive at \cite{ZhangReview}:
\begin{equation}
e^{- \frac{1}{2} \Delta \tau \hat{K}_{f} } e^{- \Delta \tau \hat{V}_{f} } e^{- \frac{1}{2} \Delta \tau \hat{K}_{f}}  = \int_{-\infty}^{\infty} d\vec{\phi} p(\vec{\phi})  \hat{B}_{f}(\vec{\phi}),
\label{eq:etauH_f_HS}
\end{equation}
where
\begin{equation}
\hat{B}_{f}(\vec{\phi}) = e^{- \frac{1}{2} \Delta \tau \hat{K}_{f} }  \left[ \prod_{i}  e^{\phi_{i} \sqrt{U_{f} \Delta \tau} \left( \hat{m}_{i\uparrow} - \hat{m}_{i\downarrow} \right)}   \right] e^{- \frac{1}{2} \Delta \tau \hat{K}_{f}}, 
\end{equation}
and the one-body term in Eq.~\ref{Vf_sqform2} can be absorbed by replacing $\epsilon_{i, \sigma}^{f}$  with $(\epsilon_{i, \sigma}^{f}+U_f/2)$ in Eq.~\ref{eq:Kf}.  

Substituting Eq.~\ref{eq:etauH_f_HS} into the expression for $Z$ in Eq.~\ref{PartitionFunction}, one arrives at the central AFQMC equation
\begin{eqnarray}
Z_f &=& \int_{-\infty}^{\infty} d\vec{\Phi} p(\vec{\Phi}) Tr \left( \hat{B}_{f}(\vec{\phi}_{l}) ... \hat{B}_{f}(\vec{\phi}_{1}) \right), 
\label{FinalPartition}
\end{eqnarray}
where $\vec{\Phi}$ denotes the full collection of auxiliary-fields at each timeslice and site and $p(\vec{\Phi})$ is the corresponding probability of selecting those fields. 

The partition function may therefore be viewed as an integral over all fields of the Gaussian probability of selecting a set of fields multiplied by the trace of single-body operators evaluated as a function of the fields. The set of fields at each timeslice and site constitutes a path in AF space. Thus, in AFQMC, one calculates the multi-dimensional partition function by stochastically sampling a set of paths in AF space and evaluating the weighted average of the trace along those paths. 

It turns out that the \emph{fermion} trace over one-body propagators can be evaluated analytically and expressed as a determinant  \cite{Hirsch}
\begin{equation}
Tr_{f} \left( \hat{B}_{f}(\vec{\phi}_{l}) ... \hat{B}_{f}(\vec{\phi}_{1}) \right) = 
Det \left[ I + B_{f}(\vec{\phi}_{l}) ... B_{f}(\vec{\phi}_{1}) \right].  
\label{eq:Tr_part}
\end{equation}
If the size of the single-particle basis (in this case the number of lattice sites) is $N$, $B_f(\vec{\phi_{k}})$ is an $N$x$N$ matrix of the propagator $\hat{B}_{f}(\vec{\phi_{k}})$ expressed in that basis, and $I$ is the corresponding unit matrix. Inserting this expression into that for the partition function, one arrives at
\begin{equation}
Z_{f} =\int_{-\infty}^{\infty} d\vec{\Phi} p(\vec{\Phi})  Det \left[ I + B_{f}(\vec{\phi}_{l}) ... B_{f}(\vec{\phi}_{1}) \right]. 
\label{eq:Z_f}
\end{equation}
In a similar vein, tracing over fermionic operators yields the fermion Green's function:
\begin{eqnarray}
G_{ij}^f 
&\equiv & \frac{ Tr_{f}\left( \hat{f}_{i} \hat{f}_{j}^{\dagger} \hat{B}_{f}(\vec{\phi}_{l}) ... \hat{B}_{f}(\vec{\phi}_{1}) \right)}{Tr_{f} \left( \hat{B}_{f}(\vec{\phi}_{l}) ... \hat{B}_{f}(\vec{\phi}_{1})  \right) } \\
&= &\bigg[\frac{I}{I+B_{f}(\vec{\phi}_{l}) ... B_{f}(\vec{\phi}_{1})}\bigg]_{ij},
\label{eq:GreensFunctionDefinition}
\end{eqnarray}
where the subscripts on the right denote the $(i,j)$th element of the matrix.
Most observables of interest may be easily expressed in terms of the single-particle Green's function using Wick's theorem \cite{Fetter}.

With Eqs. \ref{eq:Z_f} and \ref{eq:GreensFunctionDefinition} in hand, one can evaluate nearly any observable by sampling paths according to the partition function, calculating the Green's function (and hence any related observable) as a function of those paths, and weighting the resulting values by the probability of the paths sampled. We next present the formalism that allows one to do the same for bosons. 

\section{\label{sec:level1} III. Methods}

\section{\label{sec:level4} A. Finite-Temperature AFQMC for Bosons}

Following the same steps outlined for the fermion Hamiltonian, $\hat{H}_{f}$, in Section II, one can similarly derive an expression relating the boson partition function to integrals over one-body boson propagators, $B_b(\vec{\psi}_{k})$, and auxiliary fields 
$\vec{\psi}_{k}\equiv \{ \psi_{1k}, \psi_{2k}, ..., \psi_{Nk}\}$: 
\begin{eqnarray}
Z_{b} &=& Tr_{b} \left( e^{-\beta \hat{H}_{b}} \right)  \\
&=& \int_{-\infty}^{\infty} d\vec{\Phi} p(\vec{\Psi}) Tr_{b} \left( \hat{B}_b(\vec{\psi}_{l}) ... \hat{B}_{b}(\vec{\psi}_{1}) \right). \nonumber
\end{eqnarray} 
As we show in Appendix A, the trace over bosons may also be expressed as a determinant (which has been noted in other contexts before \cite{Balian, Klich, Gubernatis}):
\begin{equation}
Tr_{b} \left( \hat{B}_b(\vec{\psi}_{l})... \hat{B}_b(\vec{\psi}_{1}) \right) = Det \left[  \frac{I}{I - B_{b}(\vec{\psi}_{l})... B_{b}(\vec{\psi}_{1}) } \right], 
\label{KeyBosonRelation}
\end{equation}
allowing the partition function to be expressed as
\begin{eqnarray}
Z_{b} &=& \int_{-\infty}^{\infty} d\vec{\Psi} p(\vec{\Psi})  Det \left[  \frac{I}{I - B_{b}(\vec{\psi}_{l})... B_{b}(\vec{\psi}_{1}) } \right].  
 \label{BPartitionFunction} 
\end{eqnarray} 
Further manipulations yield the boson single-particle Green's function
\begin{eqnarray} 
G^{b}_{ij} &\equiv& \frac{ Tr\left( \hat{b}_{i} \hat{b}_{j}^{\dagger} \hat{B}_{b}(\vec{\psi}_{l})...\hat{B}_{b}(\vec{\psi}_{1})  \right) }{Tr \left(  \hat{B}_{b}(\vec{\psi}_{l}) ...\hat{B}_{b}(\vec{\psi}_{1})  \right) } \label{BosonGreensFunction} \\
&=& \bigg[\frac{I}{I-B_{b}(\vec{\psi}_{l})... B_{b}(\vec{\psi}_{1})}\bigg]_{ij}. \nonumber
\end{eqnarray}

In a boson Auxiliary-Field Quantum Monte Carlo (B-AFQMC) algorithm, one can therefore calculate  boson observables by sampling paths according to the boson partition function in Eq.~\ref{BPartitionFunction} and evaluating the weighted average of observables determined from the boson Green's function in Eq.~\ref{BosonGreensFunction}. There are only two formal differences between B-AFQMC and standard fermion AFQMC: the minus sign in front of the product of the one-body propagators, and the inverse in the determinant. These differences, however, have a large impact on how the B-AFQMC algorithm is implemented compared to standard AFQMC. As discussed in detail in Appendix B, the new form of the Green's function requires that adjustments be made to the way one stabilizes products of one-body matrices at low temperatures, while the new form of the determinant requires that adjustments be made to the way local-updates to the Green's function are computed and weights are accumulated as fields are selected at each timeslice and site. Except for these adjustments, B-AFQMC maps formally and directly onto previous AFQMC algorithms. 

\section{\label{sec:level5} B. Bose-Fermi AFQMC}

To combine AFQMC and B-AFQMC into a procedure for mixtures, one needs to decouple the Bose-Fermi coupling term in Eq. \ref{SimplifiedBFHEquation}. This can be done by reexpressing Eq.~\ref{CouplingEquation} in a 
form suitable for the HS Transformation:
\begin{eqnarray}
\hat V_{c} &=& \frac{C}{2} \sum_{i} \left[ \left( \hat{n}_{i} + \hat{m}_{i} \right)^{2} - \hat{n}_{i}^{2} - \hat{m}_{i} \right], 
\label{eq:mixV2}
\end{eqnarray} 
where for brevity we have assumed spin-polarized fermions ($\sigma=\uparrow$ only). The more general case can be handled similarly by combining the resulting fermion interaction term with $\hat{V}_{f}$. One may now apply the HS Transformation of Eq.~\ref{HSTransformEquation} to write 
each square into linear forms as we have shown in Sec.~II.B.
Note that the resulting $\hat{n}_i^2$ terms can be absorbed into the two-body boson term, $\hat V_b$, in Eq.~\ref{eq:Vb}.

An important  way to improve the efficiency of BF-AFQMC simulations is to subtract 
any background terms prior to the HS Transformation. In both boson and fermion ground-state calculations, this was shown to greatly reduce the QMC statistical fluctuations and the severity of the sign and phase problems \cite{Purwanto2, alsaidi1}.
For example, in Eq.~\ref{eq:mixV2} one would rewrite 
$ ( \hat{n}_{i} + \hat{m}_{i} )^2 \equiv \hat v^{2} $ as 
\begin{eqnarray}
\hat{v}^{2} &=&  ( \hat v- \langle \hat v\rangle)^2+2 \hat v \langle \hat v\rangle
- \langle \hat v\rangle^2,\nonumber\\
 &=&  \hat{v'}^{2} +2  \langle \hat v\rangle\,\hat v
- \langle \hat v\rangle^2
\label{eq:mfbg_sub}
\end{eqnarray} 
for each site $i$, where $\langle \hat v\rangle\equiv \langle \hat{n}_{i}  +  \hat{m}_{i} \rangle= \langle \hat{n}_{i} \rangle + \langle \hat{m}_{i} \rangle$, with $ \langle \hat{n}_{i} \rangle$ and $ \langle \hat{m}_{i} \rangle$  the average (or desired) boson 
and fermion site densities, e.g., from MFT or   
exact symmetry properties. 
The HS Transformation is then applied to $\hat {v'}^{2}$ instead 
of  $\hat {v}^{2}$, and the one-body and constant terms in Eq.~\ref{eq:mfbg_sub}
can be easily combined with other   
one-body terms in the Hamiltonian 
and absorbed into the resulting one-body propagators, $\hat B$.

The background subtraction is intimately connected with the mean-field formalism \cite{Purwanto2}. The idea is to use a form of HS Transformation to decouple $\hat {v'}^{2}$ terms 
which are zero in some mean-field framework. 
That is, setting the AF value to zero in the HS decomposition would give the corresponding 
mean-field result.
The background subtraction
is applied to all 
$\hat V_b$ and $\hat V_c$ terms; no background subtraction is applied to 
$\hat V_f$ because we have used a 
spin-decomposition (as opposed to charge) in 
Eq.~\ref{Vf_sqform2} for fermions.
The values of $\langle \hat{n}_{i} \rangle$ and $ \langle \hat{m}_{i} \rangle$ are set prior to the simulation. It should be emphasized that  
the formalism is exact independent of 
the choice of mean-field values; only the statistical errors are affected.

The combined partition function is
\begin{equation}
Z_{bf}= Tr_{b} \left[ Tr_{f} \left[ e^{-\beta \hat{H}_{bf}} \right] \right].
\end{equation}
After the HS Transformation, the fermion and boson propagators are decoupled at each timeslice and site. Because all fermion operators commute with all boson operators, the propagators may be separated into completely independent products of one-body boson and fermion propagators.
One may then evaluate the traces over these products individually to obtain
\begin{eqnarray}
&& Z_{bf} = \int_{-\infty}^{\infty} d\vec{\Psi} d\vec{\Phi} p(\vec{\Psi}, \vec{\Phi}) \label{eq:Zbf} \\
&& Det \left[  \frac{I}{I - {B}_{b}(\vec{\psi}_{l})... {B}_{b}(\vec{\psi}_{1}) } \right] Det \left[ I + {B}_{f}(\vec{\phi}_{l})... {B}_{f}(\vec{\phi}_{1})  \right], \nonumber
\end{eqnarray}	
Because Eq.~\ref{SimplifiedBFHEquation} contains three terms quadratic in the boson and fermion densities, three HS Transformations must be used at each timeslice and site to reduce these terms to one-body operators. The boson and fermion Green's functions may analogously be written as above, but with one-body matrices that now contain their respective contributions from the coupling terms. Thus, in BF-AFQMC, a generic Bose-Fermi Hamiltonian may be simulated by first rewriting all coupling terms such that they can be transformed into independent boson and fermion propagators. Once the propagators are repartitioned, the individual boson and fermion Green's functions may then be evaluated as if there was no coupling term, so long as paths are sampled from the full Bose-Fermi partition function.  

\section{\label{sec:level6} C. Importance Sampling}

Determinants are computed using a set of walkers whose weights and Green's functions are determined as each field is sampled \emph{sequentially} in imaginary time. At the beginning of our simulations, we initialize the weights, $W(\vec{\Phi}, \vec{\Psi})$, of a collection of walkers to 1. We similarly initialize each walker's Green's function to that corresponding to a trial Hamiltonian, such that

\begin{equation}
G^{b}_{ij} =\bigg[ \frac{I}{I-B^{T}_{b} ... B^{T}_{b}} \bigg]_{ij}, 
\end{equation}
and

\begin{equation}
G^{f}_{ij} = \bigg[\frac{I}{I+B^{T}_{f}...B^{T}_{f}} \bigg]_{ij}, 
\end{equation}
where $B^{T}$ is a trial one-body matrix at each timeslice. In the work that follows, the trial Hamiltonian is typically the exact Hamiltonian minus any terms quadratic in the density ($\hat v'^2$ terms, after background subtraction). Since the chemical potential corresponding to some desired filling differs between the trial and exact Hamiltonians, care must be taken to determine the appropriate chemical potential for the trial Hamiltonian before sampling proceeds so as to prevent additional statistical fluctuations. 

As each field (or fields, if multiple HS Transformations are performed) is selected at site $i$ and timeslice $k$, the weights of the walkers are multiplied by a factor, $W(\phi_{ik}, \psi_{ik})$. In the absence of importance sampling (see below), $W(\phi_{ik}, \psi_{ik})$ is the ratio of the product of the newly-updated determinants to the old determinants. Let $P^{f}_{ik}$ denote the fermion determinant constructed of fields sampled up to the $i$-th site and $k$-th timeslice
\begin{equation}
P_{ik}^{f} = Det \left[ I + \left( \prod_{m=1}^{l-k} B^{T}_{f} \right) B_{f}(\phi_{ik}...\phi_{1k})...B_{f}(\vec{\phi}_{1}) \right]
\end{equation}
and $P^{b}_{ik}$ define the corresponding boson determinant
\begin{equation}
P_{ik}^{b} = Det \left[ \frac{I}{I - \left( \prod_{m=1}^{l-k} B^{T}_{b} \right) B_{b}(\psi_{ik}...\psi_{1k})...B_{b}(\vec{\psi}_{1})} \right],
\end{equation}
where the yet unspecified AF's in the $k$-th time slice (for sites $i$ through $N$) 
can be thought of as having value zero, as mentioned in Sec.~III.B above.
Then, the weight may be defined as
\begin{eqnarray}
&&W(\phi_{ik}, \psi_{ik}) = \frac{ P^{f}_{ik} P^{b}_{ik} }{ P^{f}_{(i-1)k}   P^{b}_{(i-1)k} }.  
\label{NoImpSampling}
\end{eqnarray}
The final product of these factors over all sampled fields is proportional to the product of boson and fermion determinants for the full path that we wish to sample. As each field is sampled, the Green's functions are also updated by replacing the trial one-body matrices with the exact one-body matrices based upon the fields. The corresponding Green's function matrix, after sampling field $i$ at timeslice $k$, would therefore be
\begin{equation}
G^{b} = \frac{I}{I-B^{T}_{b}...B^{T}_{b} B_{b}( \psi_{ik}...\psi_{1k})...B_{b}(\vec{\psi}_{1})}
\end{equation}
and
\begin{equation}
G^{f} = \frac{I}{I+B^{T}_{f}...B^{T}_{f} B_{f}( \phi_{ik}...\phi_{1k})...B_{f}(\vec{\phi}_{1})}. 
\end{equation}
All trial matrices are replaced until all fields are sampled and the Green's functions correspond to those for the exact Hamiltonian. After all fields are sampled, average observables are computed. The weights and Green's functions are then reinitialized to their starting values and fields are sampled again until the desired number of samples have been collected. 

Of course, if the fields are drawn randomly according to $p(\vec{\Phi}, \vec{\Psi})$, the ratios in Eq.~\ref{NoImpSampling} will cancel in successive steps, and our sampling procedure above will be identical to simply sampling entire paths of AFs randomly and then calculating the  determinants in Eq.~\ref{eq:Zbf} as weights of the paths. The advantage of the sampling scheme above is that it allows importance sampling to be done efficiently and, as we discuss in the next subsection, constrained path and phaseless approximations to be easily  incorporated to control sign and phase problems \cite{CPMC1,ZhangReview}.

Importance sampling  uses an estimated contribution based on a trial wave function or density matrix to guide the sampling of AFs \cite{CPMC1,Purwanto1, Phaseless}. Just as gains in efficiency may be obtained by subtracting the average density from the exact density in each HS-Transformed propagator, even further gains may be obtained by subtracting a site-dependent shift, $\bar{\psi}_{i}$, from the auxiliary-field, $\psi_{i}$. This shift, called a 
force bias, effectively modifies the probability $p(\vec{\Psi})$ for sampling $\psi_{ik}$,  
to take into account the AF paths that have been built up so far, i.e., 
the prior  $\psi$ values (from $\psi_{11}$ to $\psi_{(i-1)k})$.
The shift is added by performing a change of variable in the usual HS Transformation. For example, the boson 2-body term, after absorbing the contribution from $\hat V_c$ and 
background subtraction, can be written as
\begin{eqnarray}
\label{ExampleCoefficient2}
&& e^{-\Delta \tau/2(U_{b}-C) (\hat{n}_{i} - \langle \hat{n}_{i} \rangle )^{2} } \\
&=& \frac{1}{\sqrt{2 \pi}} \int_{-\infty}^{\infty} d \psi_{i} e^{- \psi_{i}^{2}/2} e^{-\bar{\psi}_{i}^{2}/2} \nonumber \\
&& e^{\psi_{i} \bar{\psi}_{i}} e^{(\psi_{i}-\bar{\psi}_{i}) \sqrt{ -\Delta \tau (U_{b} - C) } (\hat{n}_{i} - \langle \hat{n}_{i} \rangle ) },  \nonumber \\
&=& \int_{-\infty}^{\infty} d \psi_{i} p(\psi_{i}) W^{'}(\psi_{i}, \bar{\psi}_{i}) \hat{B}(\psi_{i} - \bar{\psi}_{i})  \nonumber 
\end{eqnarray}
where shift- and field-related constants may be regrouped into an additional weighting term, $W^{'}(\psi_{i}, \bar{\psi}_{i})$, that contributes to Eq. \ref{NoImpSampling}. The one-body operator is now also a function of the shift.  

Optimal importance sampling is achieved when the shift is chosen such that the fluctuations in the weights of the walkers are minimized. At finite-temperatures (see the ground-state derivation in Purwanto and Zhang \cite{Purwanto1}), the optimal shift may be shown to be
\begin{eqnarray}
\bar{\psi}_{i} &=& - \frac{Tr \left[ \hat{v}_{i} \hat{B}(\vec{\psi}_{l}) ... \hat{B}(\vec{\psi}_{1}) \right]}{Tr \left[  \hat{B}(\vec{\psi}_{l})  ... \hat{B}(\vec{\psi}_{1}) \right] } = - \langle \hat{v}_{i} \rangle, 
\label{eq:optimal_FB}
\end{eqnarray}
where $\hat{v}_{i}$ represents the coefficient of the field in the HS-Transformed propagator. In the case of Eq.~\ref{ExampleCoefficient2},  
$\hat{v}_{i} = \sqrt{-\Delta \tau (U_{b} - C)} (\hat{n}_{i} - \langle \hat{n}_{i} \rangle )$. 
Shifts may be calculated in this way for each HS Transformation.  This importance sampling technique enables us to simulate well into the moderate-coupling regime with high-efficiency, \emph{free of any approximations}.

\section{\label{sec:level8} D. The Constrained Path and Phaseless Approximations}

As alluded to earlier, a phase problem develops whenever complex propagators produce complex determinants. When sampled by walkers, these complex determinants in turn yield complex walker weights. Although background subtraction and importance sampling, as discussed above, can help reduce statistical fluctuations, the phase problem will eventually overwhelm any simulation at sufficiently low temperatures or sufficiently large repulsive interactions. The signature of the sign or phase problem is that the weights will populate both positive and negative values on the real axis (sign problem) or arbitrary phase angles in the complex plane, resulting in dramatic cancellation and large fluctuations. The phase problem may be avoided with the phaseless approximation, an approximation that renders the weights of complex walkers real via a gauge transformation using a trial wave function or density matrix \cite{Phaseless, Purwanto1}. 

In the phaseless approximation, one first uses importance sampling as described in Section III.C to minimize the phase of the weighting factor at each step (timeslice and site).  Without importance sampling, the weighting factor is given by Eq. \ref{NoImpSampling}. With importance sampling, it becomes
\begin{eqnarray}
W(\phi_{ik}, \psi_{ik}) = \frac{ P^{f}_{ik} P^{b}_{ik}}{ P^{f}_{(i-1)k} P^{b}_{(i-1)k} } W^{'}( \phi_{ik}, \bar{\phi}_{ik}, \psi_{ik}, \bar{\psi}_{ik} ) . 
\end{eqnarray}
With the optimal choice of force bias, as we discussed in Eq.~\ref{eq:optimal_FB}, it can be shown that the overall phase accumulation is proportional to $\Delta \tau  {Im} (E_L)$, where $E_L$ is the so-called local energy \cite{Purwanto1, Phaseless}. In the case of the exact trial wave function or density matrix, the imaginary part of $E_L$ vanishes.  Once the phase is optimally reduced, the phaseless approximation omits the overall phase. It then projects the random walk to the real axis to constrain the overall phase to one gauge choice. In the finite-temperature phaseless approximation, we define the phase rotation angle $\Delta \theta$ as
\begin{equation}
\Delta \theta \equiv Im \ln \left( \frac{ P^{b}_{ik}}{ P^{b}_{(i-1)k}} \right). 
\end{equation}
The phase angle may more generally be defined in terms of the ratios of both the boson and fermion determinants, however we find that the phase problem may typically be attributed to boson fluctuations in the Hamiltonians studied here. We then multiply the modulus of the  weighting factor, $|W(\phi_{ik}, \psi_{ik})|$ by 0 if $| \Delta \theta| > \pi/2$ and $cos(\Delta \theta)$ otherwise. This keeps the walker weights real, preventing the mass cancellation of weights symptomatic of a bad phase problem. 

In addition to the phase problem from bosons, a mixture simulation may also encounter the sign problem for 
fermions at low temperatures \cite{SignProblem}, which is a special case of the phase problem. 
The phaseless approximation in the case of a real HS Transformation and real determinants 
reduces to the constrained-path approximation \cite{CPMC1}. We use this approximation to curb the sign problem in this situation. As soon as a walker's fermion determinant becomes negative, its weight is set to zero. We thus sample only those paths such that
\begin{equation}
Det \left[I + \left( \prod_{m=1}^{l-k} B^{T}_{f} \right) B_{f}(\vec{\phi}_{k})...B_{f}(\vec{\phi}_{1}) \right] > 0 
\end{equation}
for all $k$ from 0 to $l$. As previously discussed in the literature, this prevents corrupted paths whose determinants have changed sign from contributing to observable averages. 

\section{\label{sec:level1} IV. Results}

In this section,  we present illustrative results from our Bose-Fermi AFQMC method. Results are compared to those obtained from ED, MFT, and the boson worm algorithm \cite{Prokofev1, Prokofev2}. Except where indicated, our B-AFQMC and BF-AFQMC calculations were done without imposing the phaseless or constrained path approximations; some were done without importance sampling for benchmarking or testing purposes. No optimization was performed on the choice of the parameters such as the Trotter step and the intervals with which population control \cite{ZhangReview} or stabilization procedures are applied, except to ensure that the resulting bias is well within statistical errors. 

ED is a method in which \emph{exact} expectation values are calculated from eigenvalues obtained by diagonalizing the system Hamiltonian \cite{ED}. In the grand canonical ensemble, one must determine these eigenvalues for all fermion and boson particle numbers. Since a system may in principle be occupied by an infinite number of bosons, an exact ED answer would require diagonalizing an infinite number of canonical ensemble Hamiltonians. In the results that follow, we only include a truncated number of bosons sufficient to converge our results to within three decimal places. Where the system does not collapse, this is sufficient. Near collapse, however, the truncation error was visible when compared with the BF-AFQMC results and it was necessary to increase the number of bosons included in the ED. In our simple implementation, only small clusters of up to about five lattice sites could be converged to the desired filling with this accuracy. 

For larger systems for which ED fails, we compare to MFT. MFT results are expected to be accurate only in the weak-coupling regime. Nevertheless, they provide a check on our results and demonstrate for which parameters our exact approach should be particularly valuable. In our mean field calculations, we use the general Hamiltonian
\begin{eqnarray}
\hat{H}_{MF} &=& \hat{K}_b+\hat{K}_f \nonumber \\
&+& \frac{U_{b}}{2} \sum_{i} \left( 2 \hat{n}_{i} \langle \hat{n}_{i} \rangle - \hat{n}_{i} - \langle  \hat{n}_{i} \rangle^{2}  \right)  \nonumber \\
&+& U_{f} \sum_{i} \left( \langle \hat{m}_{i\downarrow} \rangle \hat{m}_{i\uparrow} + \langle \hat{m}_{i\uparrow} \rangle \hat{m}_{i\downarrow} - \langle \hat{m}_{i\uparrow} \rangle \langle \hat{m}_{i\downarrow} \rangle   \right)               \nonumber \\
&+& C \sum_{i} \left( \hat{n}_{i} \langle \hat{m}_{i} \rangle + \hat{m}_{i} \langle \hat{n}_{i} \rangle - \langle \hat{n}_{i} \rangle \langle \hat{m}_{i} \rangle \right),  
\label{GeneralBFHMFTEquation}
\end{eqnarray}
keeping only the appropriate terms for the given model. In these calculations, we self-consistently solve for the exact boson and fermion densities at each site until our answer is converged to within three decimal places. 

Outside of the weak-coupling regime, we compare our results for the Bose-Hubbard Model to those obtained from the ALPS Projects' implementation of the worm algorithm \cite{ALPS}. The worm algorithm yields exact results for bosons for any system size, in any coupling regime \cite{Prokofev1, Prokofev2}. In all of our worm calculations, we capped the number of bosons at each lattice site at a value sufficient to achieve convergence in the energies and densities.   

\section{\label{sec:level2} A. Bose-Hubbard Model}

We begin by benchmarking our results for the Bose-Hubbard model. The Bose-Hubbard model has long been the model of choice for studying condensed He-4 in porous media \cite{BoseHubbard}. It has recently been revived to model ultracold bosons in optical lattices \cite{SFMI}. The  Hamiltonian is a special case of Eq. \ref{SimplifiedBFHEquation}, with the fermion constants all set to 0. For $U_{b}<0$ in Eq.~\ref{eq:Vb} and sufficiently low temperatures and high densities, the Bose-Hubbard model is expected to exhibit collapse \cite{Purwanto1, Purwanto2}. In the examples that follow, we therefore only present results for repulsive $U_{b}$.  Our results are equally accurate for $U_{b}<0$ before the collapse point, however. Since using $U_{b}>0$ results in a phase problem, all of the results that follow are averaged over complex phases, without the phaseless approximation. The QMC results are thus expected to be exact.

As a first check, we consider a $3\times 1$ lattice, with $t_{b}=0.01$, and $\langle n_{b}\rangle=1$. In Figs.~\ref{3SiteBHVaryingUEnergies} and \ref{3SiteBHVaryingUCondFrac}, we compare our results to those from ED for the energies and condensate fractions for varying $U_b$ down to temperatures $T/t\approx.3$. Condensate fractions measure the fraction of the system lying in the lowest eigenstate of the Hamiltonian \cite{Purwanto1, Purwanto2}. As we see in both Figs.~\ref{3SiteBHVaryingUEnergies} and \ref{3SiteBHVaryingUCondFrac}, QMC is exact within error bars well beyond where the condensate fraction asymptotes to 1. This suggests that our technique can calculate correct expectation values from high temperatures corresponding to the Mott insulating regime to low temperatures corresponding to the finite-size version of a superfluid. In Fig.~\ref{3SiteBHVaryingUCondFrac}, we also plot the MFT results for the condensate fractions to illustrate the effects of fluctuations. Only one curve is shown for the MFT condensate fractions because they are independent of $U_{b}/t$. It is evident from this figure that MFT yields poor approximations to the true condensate fractions even at relatively high temperatures and low coupling strengths. Indeed, it only reproduces the exact condensate fractions throughout this limited temperature range for $U_{b}/t=.5$. As illustrated below in Fig.~\ref{2DWorm}, even in situations where mean-field condensate fractions are nearly exact, energies produced using MFT may be unreliable. This underscores the importance of using exact methods where possible. 

The data in Figs.~\ref{3SiteBHVaryingUEnergies} and \ref{3SiteBHVaryingUCondFrac} were calculated without importance sampling or the phaseless approximation. In Fig.~\ref{3SiteBHEDQMCPhaseless}, we show that we obtain the same results with improved statistics using these techniques for $U_b/t=.5$. Using importance sampling and the phaseless approximation, our error bars on the number of bosons for the same number of samples are at least halved compared to those obtained without importance sampling. Error bars on other quantities are too small to judge. In previous works, importance sampling was observed to greatly reduce the error bars in finite-temperature fermion calculations \cite{CPMC1}. Similarly, in ground state boson calculations, an order of magnitude or more improvement in efficiency is seen \cite{Purwanto1, Purwanto2}. Our phaseless calculations for finite-temperature bosons therefore do not see the dramatic error bar reductions seen in other applications. There are several reasons for this. The system size is small, such that the variations in the sampled space are much reduced compared to larger systems, where the effect of importance sampling is expected to increase significantly.
The present boson finite-temperature calculations are performed in the grand canonical ensemble, which could contribute to increased fluctuations. The main contribution to the statistical fluctuations in the boson calculations is likely from the so-called ``rogue eigenvalue" problem which we discuss in the next section.

For larger lattices, we compare to the worm algorithm. Fig.~\ref{2DWorm} demonstrates that B-AFQMC energies are consistent with worm energies for 2D systems of varying sizes for several $U_{b}$. Interestingly, as alluded to above, QMC and MFT energies differ dramatically at all but the highest of temperatures. This is even so when the energies are normalized to account for the fact that the QMC and MFT algorithms require different chemical potentials to achieve the same fixed boson number. Fig.~\ref{2DWorm} may readily be extended to larger lattices and boson-boson repulsions, but at the price of the increased sampling needed to surmount the phase problem. 

\section{\label{sec:level2} B. Spin-Polarized Bose-Fermi-Hubbard Model}

In order to illustrate our Bose-Fermi AFQMC method, we similarly apply our technique to the Bose-Fermi-Hubbard model, the standard model for studying ultracold mixture phenomenology.  As mentioned before, here, we limit ourselves to the spin-polarized Hamiltonian, namely Eq.~\ref{SimplifiedBFHEquation} with $\hat m_{i\downarrow}=0$. 

As with the Bose-Hubbard model, collapse is anticipated for $U_{b} < 0$ and any value of $C$ for densities sufficiently large that the boson-boson attraction term dominates the linear coupling term. If $U_{b}=0$ and the boson-boson interaction does not dominate, collapse may also be observed for a sufficiently large and negative $C$. The phase problem is observed whenever $C > 0$ or $U_{b} > C$. We thus again simulate amidst the phase problem so as to at once avoid collapse and demonstrate the accuracy of our algorithm despite complex phases. 

As our first example, we consider a 2-site Bose-Fermi-Hubbard model with varying $U_{b}=C$, $t_{b}=t_{f}=0.01$, and $\langle n_{b} \rangle=\langle n_{f\uparrow}\rangle=1$. We find that our results for the potential energies, kinetic energies, condensate fractions, and double occupancies per site \cite{DoubleOccupancy} agree with ED to within small error bars for $U_{b}/t=C/t$ values up to 13. $U_{b}/t=C/t$ ratios up to 7 are shown in Fig.~\ref{2SiteBFHAll} for the sake of clarity. These results demonstrate the correctness of our algorithm and implementation, and that exact computations are feasible for moderate coupling strengths amidst an appreciable phase problem. We expect that our ability to calculate observables amidst such large phase problems will diminish with larger system sizes where fewer samples may be taken within a fixed time. More sophisticated sampling techniques, better handling of the ``rogue eigenvalue problem" (see below), and the use of the phaseless approximation will drastically improve the statistical accuracy.   

Lastly, as a check on our mixture algorithm for larger systems sizes, we compare to results from MFT in the limit of small $U_{b}$ and $C$. Our results in Fig.~\ref{2DBFMFTQMC} are in concurrence with those from MFT for up to 8x8 systems (larger sizes are not pictured here). A similar comparison, not presented here, was made for the Bose-Hubbard Hamiltonian and yielded analogous results. In both cases, MFT results compare well with QMC results until the two begin to deviate at lower temperatures, as expected. Because there are a limited number of exact methods for multidimensional mixtures to which we can compare, we reserve further mixture examples and applications for a future publication. 

\section{\label{sec:level1} V. Discussion}

\section{\label{sec:level2} A. Challenges}

As the results presented in this work demonstrate, our algorithm represents, in principle, an exact method for simulating the thermodynamic behavior of an essentially arbitrary lattice system composed of interacting bosons and fermions. Nevertheless, its performance is still hindered by several practical challenges. 

One of the more benign challenges relates to the estimation of the correct chemical potentials for desired fillings. In order to simulate a mixture with the desired fillings in the grand canonical ensemble, one must estimate not only the correct fermion chemical potential, but the correct boson chemical potential as well. This task is particularly laborious for bosons since their fillings may change especially rapidly with chemical potential. When fillings change more gradually with chemical potential, such as in the Mott insulator or normal liquid regimes, iterative methods may be employed. Outside of such regimes, particularly near or in superfluid phases, such methods fail because incorrect or unphysical chemical potentials may yield seemingly correct fillings within error bars. 

A second challenge to our algorithm is posed by the phase problem. As discussed in Section III.D, whenever propagators become complex, walker weights and Green's functions acquire a complex phase. When this phase grows particularly large, controlling statistical fluctuations becomes a computational challenge. The severity of the phase problem depends upon the model and simulation parameters. For the Bose-Hubbard model, the phase problem develops for positive $U_{b}$; for the Bose-Fermi-Hubbard model it is present whenever $C > 0$ or $U_{b} > C$.  As with the related sign problem in fermion QMC, the severity of the phase problem grows exponentially with system size or inverse
temperature. This means that for large systems and at low temperatures, we need 
to properly impose constraints that systematically bias the results. The performance of the constraint in ground
state calculations should provide a ``lower bound'' to the quality of the approximation in 
these finite-temperature calculations. As was previously discussed, importance sampling can significantly reduce statistical fluctuations, and where importance sampling fails, the phaseless approximation may be invoked. However, how the approximation performs across a phase transition, especially when the constraining trial density matrix is poor, remains to be studied. 

Perhaps the biggest issue in the present formulation relates to the fact that in the grand canonical ensemble boson numbers may fluctuate in an unbounded manner.
In the auxiliary-field formalism, the many-body problem is turned into multiple independent-particle problems in external fields.  By fluctuation of the external fields, the target 
chemical potential may be too high for a particular independent-particle path, which would 
result in a condensate with an infinite number of particles. We have termed this the ``rogue eigenvalue problem.'' 

As seen in Eq. \ref{KeyBosonRelation}, our boson partition function is expressed as a determinant of a matrix whose denominator may approach or fall below zero. This happens whenever the largest eigenvalue of the product of one-body boson matrices approaches or surpasses unity. Although it is unphysical for the leading eigenvalue to surpass one - and indeed, it never does in our completely deterministic mean-field calculations - our walkers may stochastically sample such unphysical paths and their related ``rogue eigenvalues.'' Walkers whose eigenvalues have surpassed one at any point in imaginary time possess corrupted paths that develop appreciable phase problems more severe than those seen in fermion systems and unique to simulations of bosons in the grand canonical ensemble. This is the leading challenge which impacts the effectiveness of the algorithm even in the presence of importance sampling and the phaseless approximation. In order to obtain sensible results well into condensed phases where eigenvalues may approach one on physical grounds, we must therefore prevent walkers from sampling rogue paths. One facile method for suppressing rogue paths used to produce many of the figures in this paper involved using larger $\langle \hat{v}_{i} \rangle$ values than the mean-field values. Instead of setting $\langle \hat{v}_{i} \rangle$ in Eq.~\ref{eq:mfbg_sub} to the sum of the mean-field densities at a given site, we set it to larger values that increase the effective chemical potential seen by the Green's functions. This reduces the risk of a rogue eigenvalue problem at the cost of increased phase fluctuations, which can be surmounted by increased averaging. Further details about this approach and more sophisticated ones will be presented in an upcoming publication. 

\section{\label{sec:level2} B. Conclusions}
In this work, we have outlined a new algorithm that enables the exact calculation of the thermodynamic properties of BF mixtures in multiple dimensions over a wide range of parameters. This algorithm enables us to sample the boson partition function and calculate boson expectation values much as one would sample the fermion partition function and calculate fermion expectation values using conventional fermion AFQMC. Our method is, in principle, exact and we have demonstrated its accuracy by comparing our results to those obtained via ED and MFT for the Bose-Hubbard and spin-polarized Bose-Fermi Hubbard models. Approximations need only be invoked when stochastic errors stemming from the sign and phase problems become uncontrollable. Because our algorithm is at once exact and computationally tractable, we believe it is uniquely positioned to answer many open questions about the Bose-Fermi phase diagram and recent mixture experiments. Our algorithm is particularly well-suited for the study of inhomogenous phases with long-range correlations, which cannot be reliably captured by mean-field approaches. We leave applications of our method to problems of genuine physical interest to future publications. 

\section{Acknowledgements}
B.M.R. thanks James Gubernatis, Emanuel Gull, and Andrew Millis for enlightening discussions. B.M.R.'s contributions were supported by the DOE Computational Science Graduate Fellowship (DE-FG02-97ER-25308) and the NSF Graduate Research Fellowship Program. S.Z. was supported by ARO (56693-PH) and NSF (DMR-1006217). 

\section{\label{sec:appendixA} Appendix A: Derivation of Boson Partition and Green's Functions}

In this Appendix, we derive expressions for the boson partition and Green's functions that are essential to our boson and Bose-Fermi mixture AFQMC algorithms. These expressions have appeared in other contexts elsewhere \cite{Balian, Klich, Gubernatis}. We derive these in detail below, drawing from Refs.~\cite{Hirsch} and \cite{HamannFahy}.

The fundamental relationship we aim to prove relates the trace of a product of one-body operators to a determinant  
\begin{equation}
Tr_{b} \left[ e^{- b_{i}^{\dagger} A_{ij} b_{j} } e^{-b_{i}^{\dagger} B_{ij} b_{j} } \right] = Det \left[ \frac{I}{I-e^{-A}e^{-B}} \right], 
\label{KeyRelationshipTrace}
\end{equation}
where $b_{i}^{\dagger}, b_{i}$ are boson creation and annihilation operators at site $i$ and $A$ and $B$ are arbitrary matrices of coefficients. 
Let us use $\hat b^\dagger$ to denote a {\em row\/} vector of boson creation operators:
\begin{equation}
\hat b^\dagger \equiv \{b^\dagger_{1},  b^\dagger_{2}, \cdots, b^\dagger_{N}\},
\end{equation}
where $N$ is the size of the one-particle basis. Correspondingly, $\hat b$ 
will denote a {\em column} vector of annihilation operators. A general one-body operator 
$\hat A$ is then
\begin{equation}
\hat A = \hat b^\dagger {\rm A}\hat b= \sum_{ij} b_{i}^{\dagger} A_{ij} b_{j},  
\end{equation}
which is a scalar and is defined by the matrix ${\rm A}$ whose matrix elements are given by $A_{ij}$. 

To prove Eq.~\ref{KeyRelationshipTrace}, we first prove the following identity
\begin{equation}
e^{-\hat A} e^{-\hat B}=e^{-\hat C},
\label{KeyRelationship1B_Op}
\end{equation}
where the matrix ${\rm C}$ defining the one-body operator $\hat C$ is given by
$e^{- {\rm C}} \equiv e^{- {\rm A}} e^{- {\rm B}}$.
Once Eq.~\ref{KeyRelationship1B_Op} is proven, we can easily go to the diagonal basis
to obtain Eq.~\ref{KeyRelationshipTrace}.
Let $U^{\dagger} {\rm C} U= Diag[ c_{i} ]$, where $c_{i}$ are the eigenvalues of the matrix ${\rm C}$, and $\hat b'_{i}=U_{ij}^{\dagger}b_{j}$. Then, 
\begin{eqnarray}
Tr_{b} \left[ e^{ -b_{i}^{\dagger} C_{ij} b_{j} } \right] &=& Tr_{b} \left[ e^{-\sum_{i} \hat b'^\dagger_{i} c_{i} \hat{b'}_{i} } \right] \nonumber \\
&=& \prod_{i} \sum_{n_{i}=0} ^\infty e^{-n_{i} c_{i}} \nonumber \\
&=& \prod_{i} \left[1-e^{-c_{i}} \right]^{-1} \nonumber \\
&=& Det \left[ \left[ I - e^{-{\rm C}} \right]^{-1} \right]. 
\label{CijTrace}
\end{eqnarray}

To prove Eq.~\ref{KeyRelationship1B_Op}, we consider the operation $\hat A\hat b^\dagger$.
Using the boson commutation relation: $b_j b^\dagger_k=\delta_{jk}+b^\dagger_k b_j$,
we have
 \begin{equation}
\hat A   b^\dagger_k = \sum_{ij}  b_{i}^{\dagger} A_{ij} b_{j} \,b^\dagger_k=\sum_{i}   b_{i}^{\dagger} A_{ik} +b^\dagger_k\sum_{ij}  b_{i}^{\dagger} A_{ij} b_{j},
\end{equation}
which gives
\begin{equation}
\hat A   \hat b^\dagger = \hat b^\dagger\cdot ({\rm A}+{\rm I} \hat A),
\label{Tonb}
\end{equation}
where ${\rm I}$ is an $N\times N$ unit matrix. Note the left-hand side is a scalar times a row vector
while the right-hand side is a row vector times a matrix. Repeated application of this equation yields
\begin{equation}
\hat A^m   \hat b^\dagger = \hat b^\dagger\cdot ({\rm A}+{\rm I} \hat A)^m,
\label{Tonb_m}
\end{equation}
for any positive integer $m$. Thus
\begin{equation}
e^{-\hat A} \,\hat b^\dagger = \hat b^\dagger\cdot e^{-({\rm A}+{\rm I} \hat A)}=\hat b^\dagger\cdot e^{-{\rm A}}\,e^{- \hat A},
\label{expTonb}
\end{equation}
where in the last step the exponential can be broken up as the two parts commute. 
This is similar to the equation for fermions \cite{HamannFahy}. 

Now we consider an arbitrary  
single-boson state 
\begin{equation}
| \phi \rangle \equiv \hat \phi^\dagger|0\rangle\equiv \hat b^\dagger \cdot \phi|0\rangle=\sum_n \phi_n b_n^\dagger\,|0\rangle, 
\end{equation}
where $\phi$ is a {\em column} vector containing the orbital coefficients $\phi_i$.
The operation of the one-body propagator $e^{-\hat A}$ on the state leads to
\begin{equation}
e^{-\hat A}\,|\phi\rangle=e^{-\hat A}\, \hat b^\dagger \cdot \phi|0\rangle
=\hat b^\dagger\cdot e^{-{\rm A}}\cdot\phi \,|0\rangle, 
\end{equation}
where in the last step we have used the fact  $e^{-\hat A}|0\rangle=|0\rangle$.
Similarly, for a two-boson state
\begin{equation}
| \psi,\phi \rangle \equiv  \hat \psi^\dagger \hat \phi^\dagger\,|0\rangle= 
(\hat b^\dagger \cdot \psi) (\hat b^\dagger \cdot \phi)|0\rangle,
\end{equation}
we have
\begin{equation}
e^{-\hat A}\,|\psi,\phi\rangle
=(\hat b^\dagger\cdot e^{-{\rm A}}\cdot\psi)  (\hat b^\dagger\cdot e^{-{\rm A}}\cdot\phi) \,|0\rangle.
\end{equation}
Proceeding inductively, we see that the effect of any single-particle propagator $e^{-\hat A}$ on 
any $n$-particle state (including states in which some orbitals are identical, i.e., multiple
bosons occupying the same 1-particle orbital) is simply to modify each orbital by the matrix  $e^{-{\rm A}}$. Applying this twice leads to the proof of Eq.~\ref{KeyRelationship1B_Op}.

With an expression for the trace in hand, we can evaluate the related boson Green's function. The Green's function may be written as
\begin{eqnarray}
G^{b}_{ij} &=& \frac{ Tr_{b} \left[  b_{i} b_{j}^{\dagger}  e^{-\hat B} e^{-\hat A} \right]}{Tr_{b} \left[ e^{-\hat B} e^{-\hat A} \right]}
= \frac{ Tr_{b} \left[ b_{i} b_{j}^{\dagger} e^{-\hat C} \right]}{Tr_{b} \left[ e^{-\hat C}  \right]},
\end{eqnarray}
where we have used $e^{-\hat A}$ and  $e^{-\hat B}$ to represent the product of one-boson 
propagators for the time slices $m\le k$ and $m>k$, respectively, with the equal-time Green's function measured at time-slice $k$, and $e^{-\hat C}= e^{-\hat A}e^{-\hat B}$. Transforming to the one-particle basis $\{ |\nu\rangle \}$ that diagonalizes $\hat C$, as in Eq.~\ref{CijTrace}, we obtain:
\begin{eqnarray}
G^{b}_{ij} &=& \frac{ Tr_{b} \left[ ( \delta_{ij} + b_{j}^{\dagger}  b_{i} ) \prod_{\nu} e^{- {\hat b}_{\nu}^{\dagger} c_{\nu} {\hat b}_{\nu}} \right]} { 
Tr_{b} \prod_{\nu} ^{- {\hat b}_{\nu}^{\dagger} c_{\nu} {\hat b}_{\nu}} } \nonumber \\
&=& \delta_{ij} + \sum_{\nu'} \langle \nu' | j\rangle \langle i | \nu' \rangle \frac{ Tr_{b} \left[ b_{\nu'}^{\dagger} {\hat b}_{\nu'} \prod_{\nu} e^{-{\hat b}_{\nu}^{\dagger} c_{\nu} {\hat b}_{\nu} } \right] }
{ Tr_{b} \prod_{\nu} ^{- {\hat b}_{\nu}^{\dagger} c_{\nu} {\hat b}_{\nu}} } \nonumber \\
&=&  \delta_{ij}  - \sum_{\nu'} \langle \nu' | j \rangle \langle i | \nu' \rangle \frac{d}{d c_{\nu'}} \ln Tr_{b} \left[ \prod_{\nu'} e^{-{\hat b}_{\nu'}^{\dagger} c_{\nu'} {\hat b}_{\nu'}} \right] \nonumber \\
&=&  \delta_{ij} +  \langle i | \left[ \sum_{\nu'} |\nu' \rangle \frac{e^{- c_{\nu'}}}{1- e^{-c_{\nu'}}} \langle \nu' |\right] |j \rangle \nonumber \\
&=&  \left[\frac{I}{I - e^{-\hat C}}\right]_{ij}. 
\label{eq:GrFcnDer}
\end{eqnarray}
In equilibrium AFQMC simulations, $e^{-\hat{C}}$ represents the decomposition of the density matrix $e^{-\beta \hat{H}}$ as the product of time-sliced exponentials of quadratic operators, $\hat{B}(\vec{\phi}_{l})...\hat{B}(\vec{\phi}_{1})$, with the corresponding time-ordering as defined by $k$,
where the Green's function is measured. With these equations, one can readily extend fermion AFQMC techniques to bosons. 

\section{\label{sec:appendixB} Appendix B: Algorithmic Details for Working with Boson Green's Functions}

The form of the boson Green's function necessitates three changes to the usual fermion AFQMC algorithm. The first two changes pertain to the equations for calculating the ratio of determinants and the updated boson Green's function after each selection of a new field. The last pertains to the computational stability and conditioning of boson Green's functions at low temperatures. 

While the boson Green's function may be recalculated from scratch each time it is altered, it is numerically cheaper to use the Sherman-Morrison-Woodbury formula, which yields the inverse of an invertible matrix plus a dyadic product. The formulas for performing rank-one updates on the fermion Green's function are well-known \cite{Stabilization, Bai}. Following Bai's derivation for fermions \cite{Bai}, here we derive the related formulas for boson Green's functions,
$I/(I-e^{-\hat C})$, as given in Eq.~\ref{eq:GrFcnDer}.

Let $M_{1}$ be the inverse of a boson Green's function before the selection of a field and $M_{2}$ be that after the selection of a field. From Eq.~\ref{eq:GrFcnDer}, we can write these as 
\begin{equation}
M_{1} = I - F V_{1}
\end{equation}
and 
\begin{equation}
M_{2} = I - F V_{2}. 
\end{equation}
$F$ represents a matrix appropriate for the corresponding $\hat C$. $V_{1}$ and $V_{2}$ are diagonal matrices, only differing at the $i^{\rm th}$ element. With no loss of generality, 
let us assume $i=1$. Then
\begin{equation}
V_{1}^{-1} V_{2} = I + \alpha e_{1} e_{1}^{T}, 
\end{equation}
where
\begin{equation}
\alpha \equiv \frac{V_{2}(1, 1)}{V_{1}(1, 1)}-1. 
\end{equation}
As usual, $e_{1}$ represents the first column of the identity matrix. $M_{2}$ may then be reexpressed in terms of $M_{1}$
\COMMENTED{
\begin{eqnarray}
M_{2} &=& I - FV_{1} - FV_{1}(V_{1}^{-1} V_{2} - I) \nonumber \\
&=& M_{1} - FV_{1}(V_{1}^{-1}V_{2} - I) \nonumber \\
&=& M_{1} - \alpha F V_{1} e_{1} e_{1}^{T} \nonumber \\
&=& M_{1} + \alpha (M_{1} - I) e_{1} e_{1}^{T} \nonumber \\
&=& M_{1} \left[ I + \alpha (I-M_{1}^{-1}) e_{1} e_{1}^{T} \right].
\label{M2Equation}
\end{eqnarray}
}
\begin{eqnarray}
M_{2} &=& I - FV_{1} - FV_{1}(V_{1}^{-1} V_{2} - I) \nonumber \\
&=& M_{1} - \alpha F V_{1} e_{1} e_{1}^{T} \nonumber \\
&=& M_{1} \left[ I + \alpha (I-M_{1}^{-1}) e_{1} e_{1}^{T} \right].
\label{M2Equation}
\end{eqnarray}
Expressing $M_{2}$ in terms of $M_{1}$ in this form allows one to readily determine the ratio of determinants, $r^{b}$, of the respective matrices. As discussed in Section III.C, $r^{b}$ must be included in the weighting factor that multiplies the overall walker weight after each field selection. For bosons, the ratio of interest is
\begin{eqnarray}
r^{b} \equiv \frac{ Det[I/M_{2}] }{ Det[I/M_{1}]} = \frac{ Det[M_{1}]}{Det[M_{2}]}. 
\end{eqnarray}
From above, we have
\begin{eqnarray}
1/r^{b} &=& Det[M_{2}]/Det[M_{1}] \nonumber\\
&=& Det[  I + \alpha (I-M_{1}^{-1}) e_{1} e_{1}^{T}  ] \nonumber \\
&=& 1 + \alpha ( 1 - e_{1}^{T} M_{1}^{-1} e_{1} ).
\end{eqnarray}
Thus, 
\begin{equation}
r^{b} = \frac{1}{1 + \alpha ( 1 - e_{1}^{T} M_{1}^{-1} e_{1} )}. 
\end{equation}
If one were to sample boson determinants using the Metropolis algorithm, it is $r^{b}$ that would be used in the acceptance criterion. 

The updated Green's function may furthermore be obtained by inverting Eq. \ref{M2Equation}.  Taking the inverse, we have
\begin{equation}
M_{2}^{-1} = \left[ I + \alpha (I-M_{1}^{-1}) e_{1} e_{1}^{T} \right]^{-1} M_{1}^{-1}. 
\end{equation}
Using the Sherman-Morrison-Woodbury formula,
\begin{equation}
(A+u v^{T})^{-1} = A^{-1} - \frac{ A^{-1} u v^{T} A^{-1}}{1+v^{T} A^{-1} u},
\end{equation}
and letting $A=I$, $u=\alpha (I-M_{1}^{-1})e_{1}$, and $v^{T}=e_{1}^{T}$, we then have
\begin{eqnarray}
M_{2}^{-1} &=& \left[ I - \frac{ \alpha (I-M_{1}^{-1}) e_{1} e_{1}^{T} }{1+\alpha e_{1}^{T} (I-M_{1}^{-1}) e_{1} } \right] M_{1}^{-1} \nonumber \\
&=& M_{1}^{-1} - \frac{\alpha}{r^{b}} (I-M_{1}^{-1}) e_{1} e_{1}^{T} M_{1}^{-1}.  
\end{eqnarray}
Since $M_{1}^{-1}$ is simply the previous boson Green's function and $\alpha$ and $r^{b}$ have been calculated, this equation represents a facile way of updating the boson Green's function. Analogous equations may be derived for other diagonal sites.    

In addition to these adjustments to the local updating scheme, a slight change must also be made to the way one inverts the boson Green's function. Just as special care must be taken to invert the ill-conditioned denominator of the fermion Green's function at low temperatures, care must similarly be taken to invert the denominator of the boson Green's function. One should therefore perform the same $UDV$-decomposition used for fermions \cite{Stabilization} on bosons, but with a sign change reflecting the opposite sign that appears in the denominator of the boson Green's function: 
\COMMENTED{
Therefore, whereas one would factor the fermion Green's function into

\begin{eqnarray}
G^{f} &=& [ I + UDV]^{-1} = V^{-1} [U^{-1} V^{-1} + D]^{-1} U^{-1} \nonumber \\
&=& V^{-1} [U^{'} D^{'} V^{'} ]^{-1} U^{-1}, 
\end{eqnarray}
one should factor the boson Green's function into
}
\begin{eqnarray}
G^{b} &=& [ I - UDV]^{-1} = V^{-1} [U^{-1} V^{-1} - D]^{-1} U^{-1} \nonumber \\
&=& V^{-1} [U^{'} D^{'} V^{'} ]^{-1} U^{-1}. 
\end{eqnarray}
In the above, $U, U^{'}$ are orthonormal matrices, $D, D'$ are diagonal matrices, and $V, V'$ are upper-triangular.

\begin{widetext}

\section{Figures}
\begin{figure}
\begin{center}\includegraphics[width=160mm]{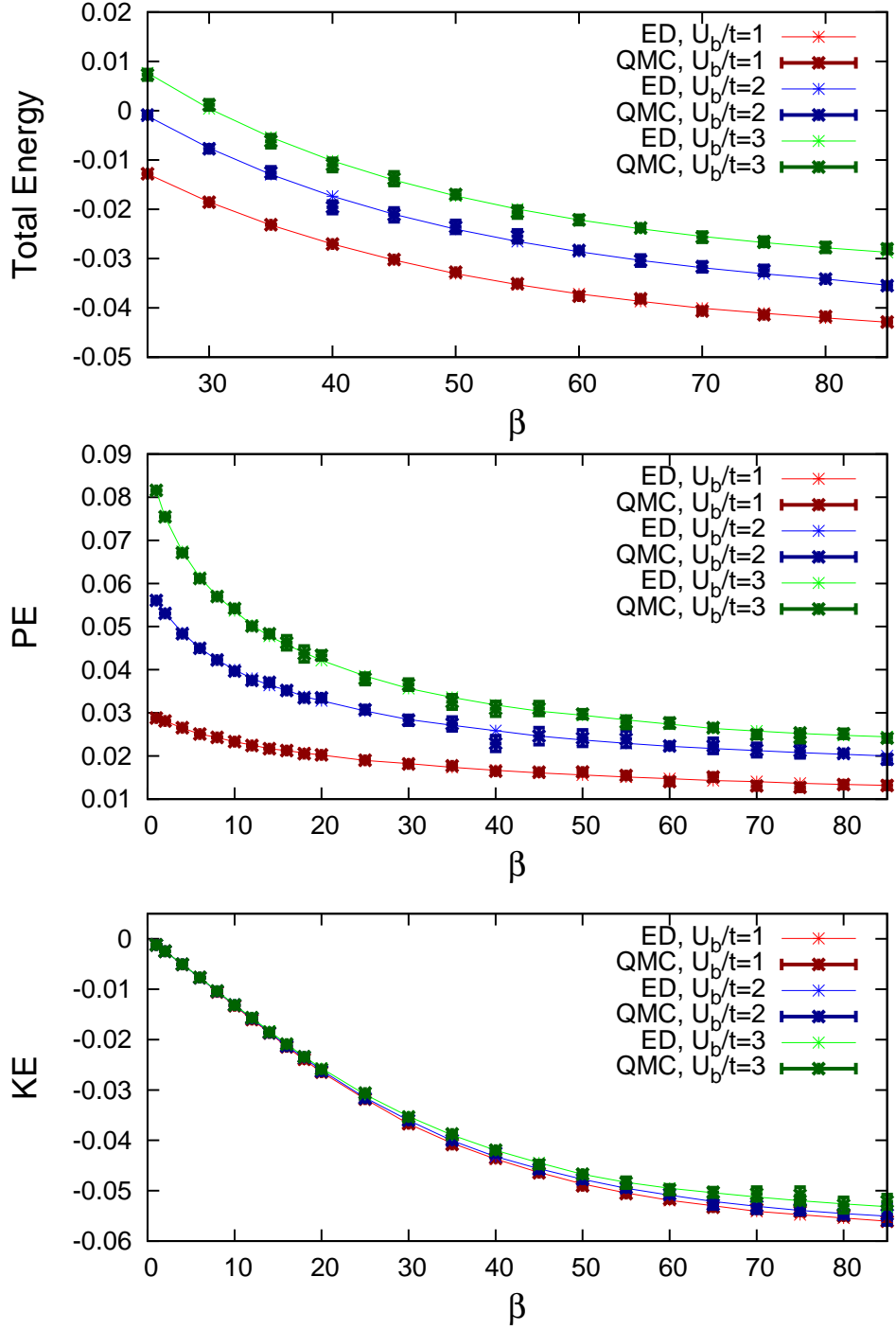}\end{center}
\caption{The total, kinetic (KE), and potential (PE) energies of a 3-site Bose-Hubbard Model simulated for several values of $U_{b}$, $t_{b}=0.01$, and $\langle n_{b} \rangle=1$ using both ED and QMC. $\beta$ denotes the inverse temperature. Agreement is within error bars for all points depicted.}
\label{3SiteBHVaryingUEnergies}
\end{figure} 

\begin{figure}
\begin{center}\includegraphics[width=160mm]{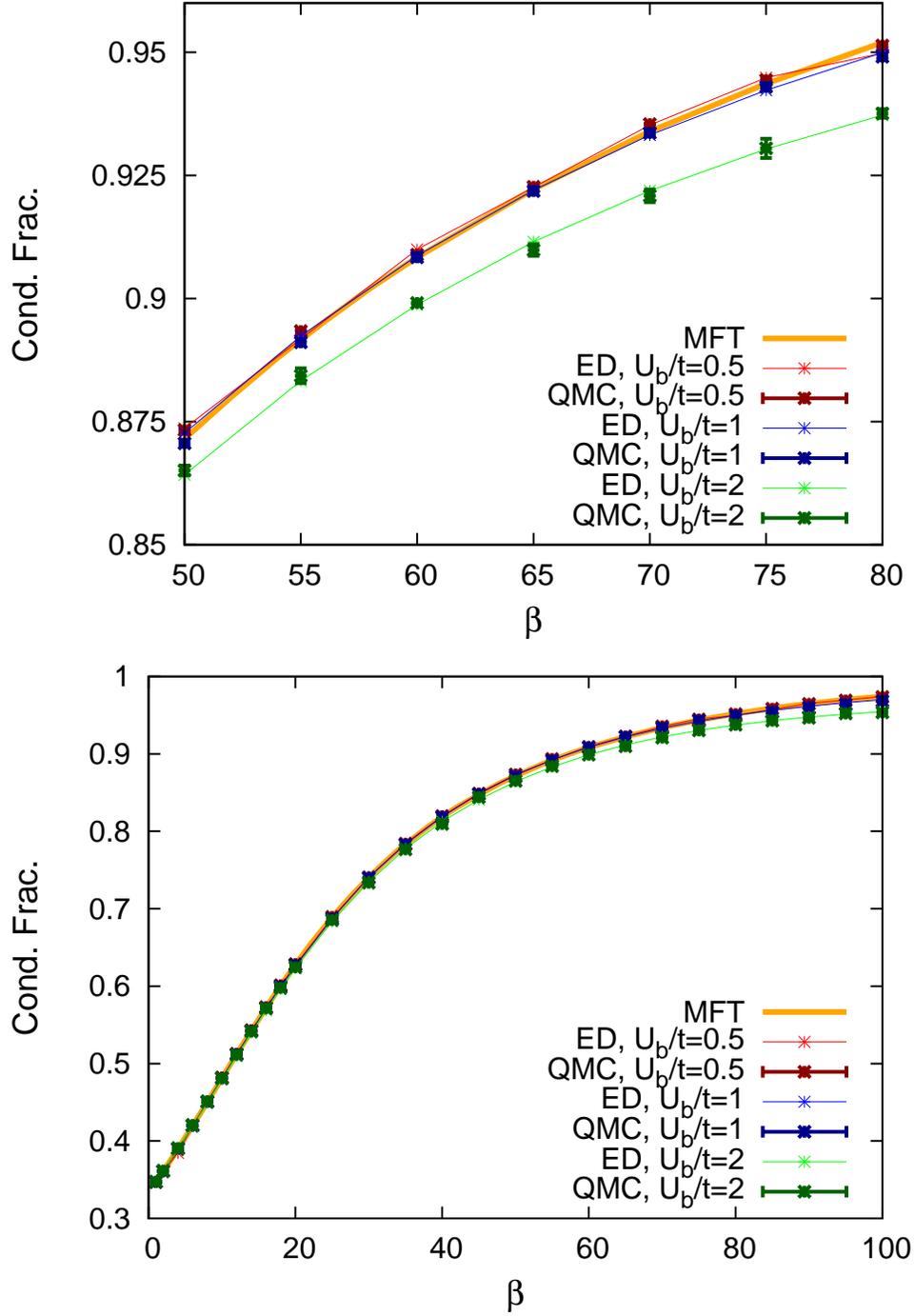}\end{center}
\caption{3-Site Bose-Hubbard Model simulated for several values of $U_{b}$, $t_{b}=0.01$, and $\langle n_{b} \rangle=1$ using ED, QMC, and MFT. Because MFT yields the same non-interacting value of the condensate fraction regardless of $U_{b}$, only one mean-field curve is shown above. $\beta$ denotes the inverse temperature. Agreement between ED and QMC is exact within error bars. MFT is only accurate for small $U_{b}/t$.}
\label{3SiteBHVaryingUCondFrac}
\end{figure} 

\begin{figure}
\begin{center}\includegraphics[width=160mm]{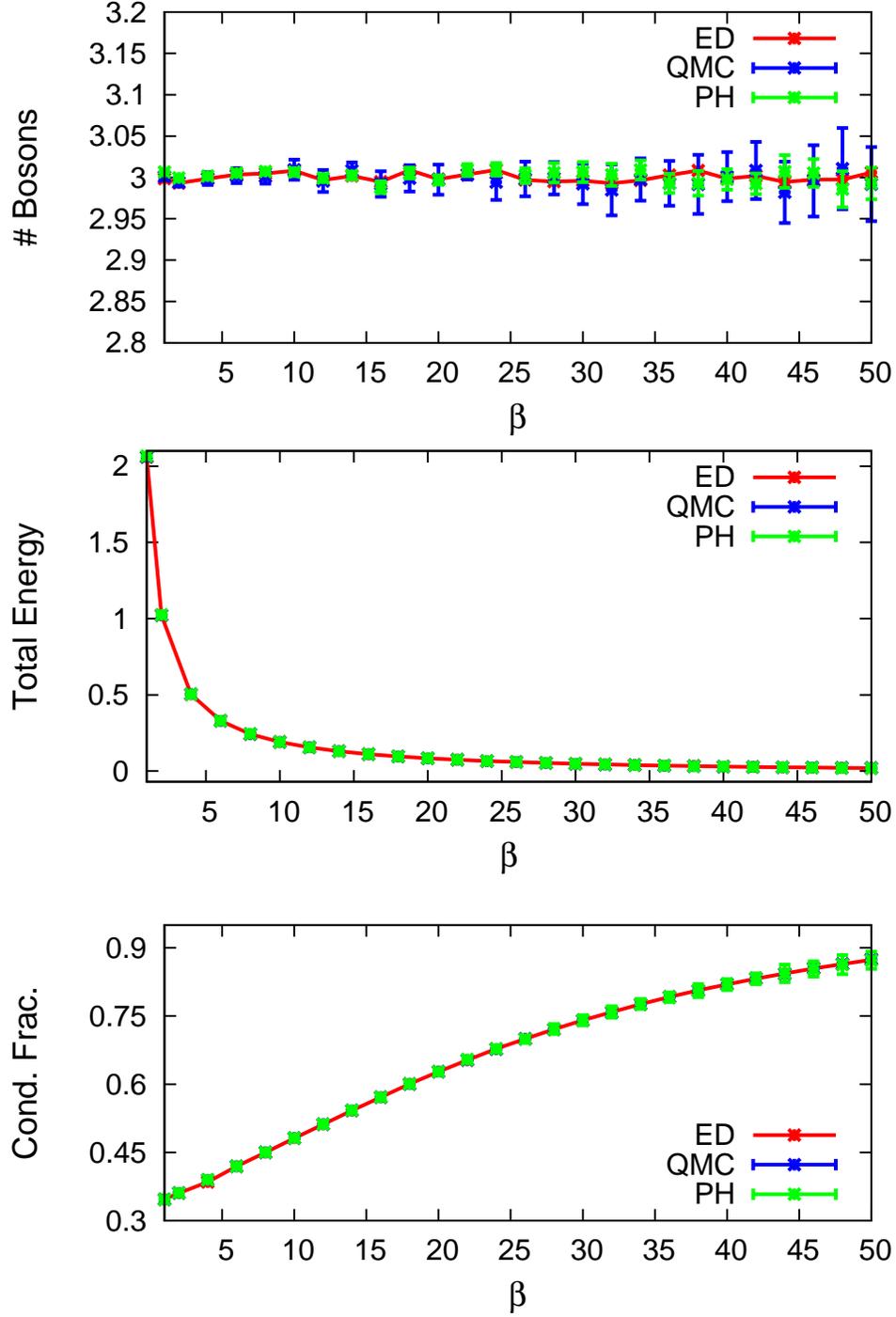}\end{center}
\caption{Number of bosons, total energies, and condensate fractions using ED, exact QMC, and the phaseless (PH) approximation for a 3-site Bose-Hubbard model with $U_{b}/t=0.5$, $t_{b}=0.01$, and $\langle n_{b} \rangle = 1$. $\beta$ denotes the inverse temperature. All points were produced with a timeslice of $\Delta \tau=.025$ and 50000 samples. The phaseless approximation reduces the size of the error bars on the number of bosons by at least half with respect to the exact QMC error bars.}
\label{3SiteBHEDQMCPhaseless}
\end{figure} 

\begin{figure}
\begin{center}\includegraphics[width=160mm]{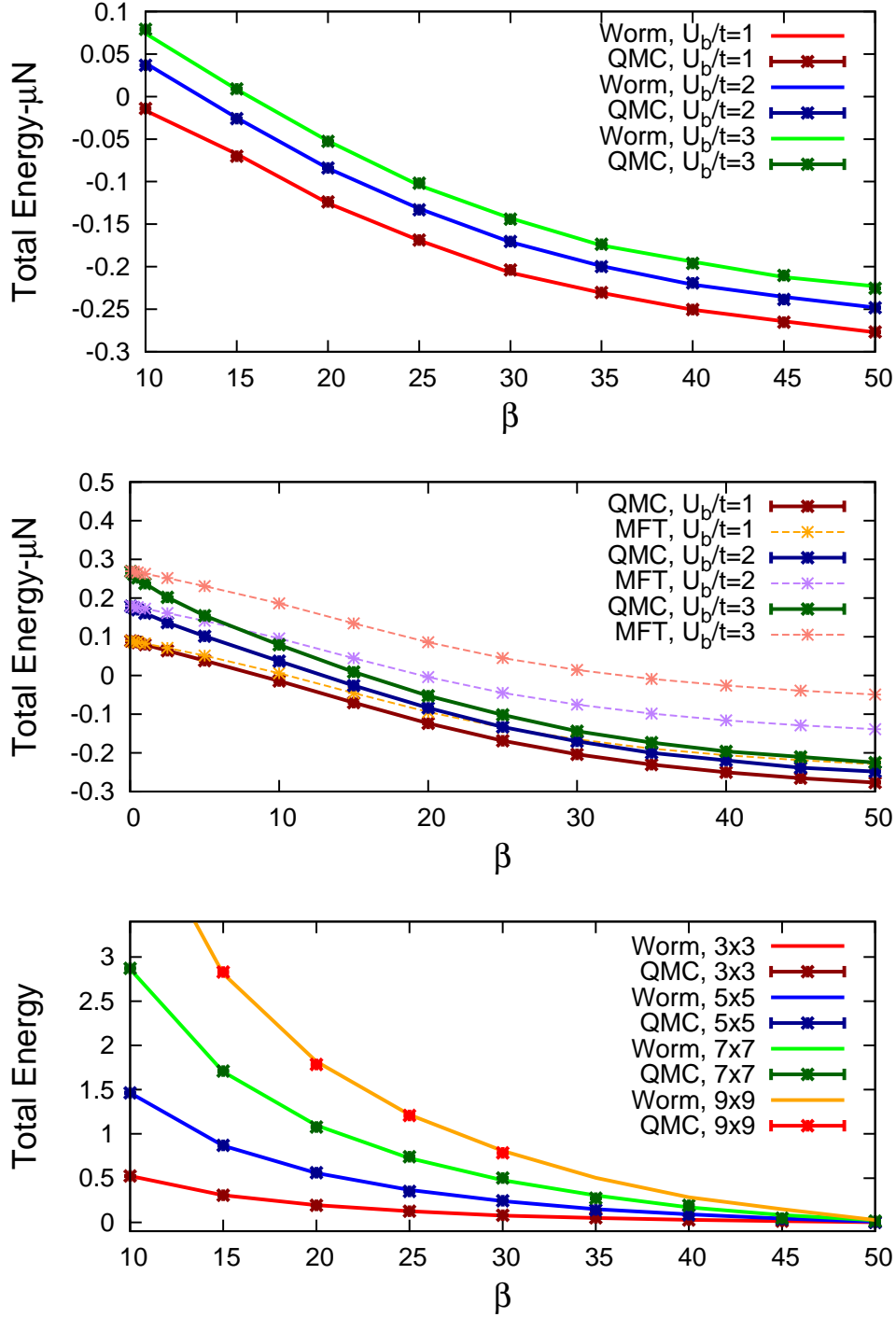}\end{center}
\caption{QMC vs. worm algorithm total energies for 2D Bose-Hubbard models with $t_{b}=0.01$ and $\langle n_{b} \rangle=1$. $\beta$ denotes the inverse temperature. Top: Total energies minus chemical potential contributions from the worm and B-AFQMC algorithms with decreasing temperature for a 3x3 Bose-Hubbard model for several $U_{b}$. Center: Total energies minus chemical potential contributions from B-AFQMC and MFT with decreasing temperature for a 3x3 Bose-Hubbard model for several $U_{b}$ (note different scales on the horizontal axis). The QMC data is the same as used in the top panel. Bottom: Total energies with decreasing temperature for 2D models of varying size for $U_{b}/t=0.5$. Total energy minus chemical potential contributions is plotted above in order to remove any discrepancies resulting from the fact that B-AFQMC and MFT require different chemical potentials to achieve the same boson densities. B-AFQMC can accurately reproduce energies for varying systems sizes and interaction strengths as seen by comparing to the worm algorithm. The B-AFQMC's reach is only limited by the phase problem. Worm and B-AFQMC energies dramatically differ from those obtained using MFT at lower temperatures.}
\label{2DWorm}
\end{figure}

\begin{figure}
\begin{center}\includegraphics[width=160mm]{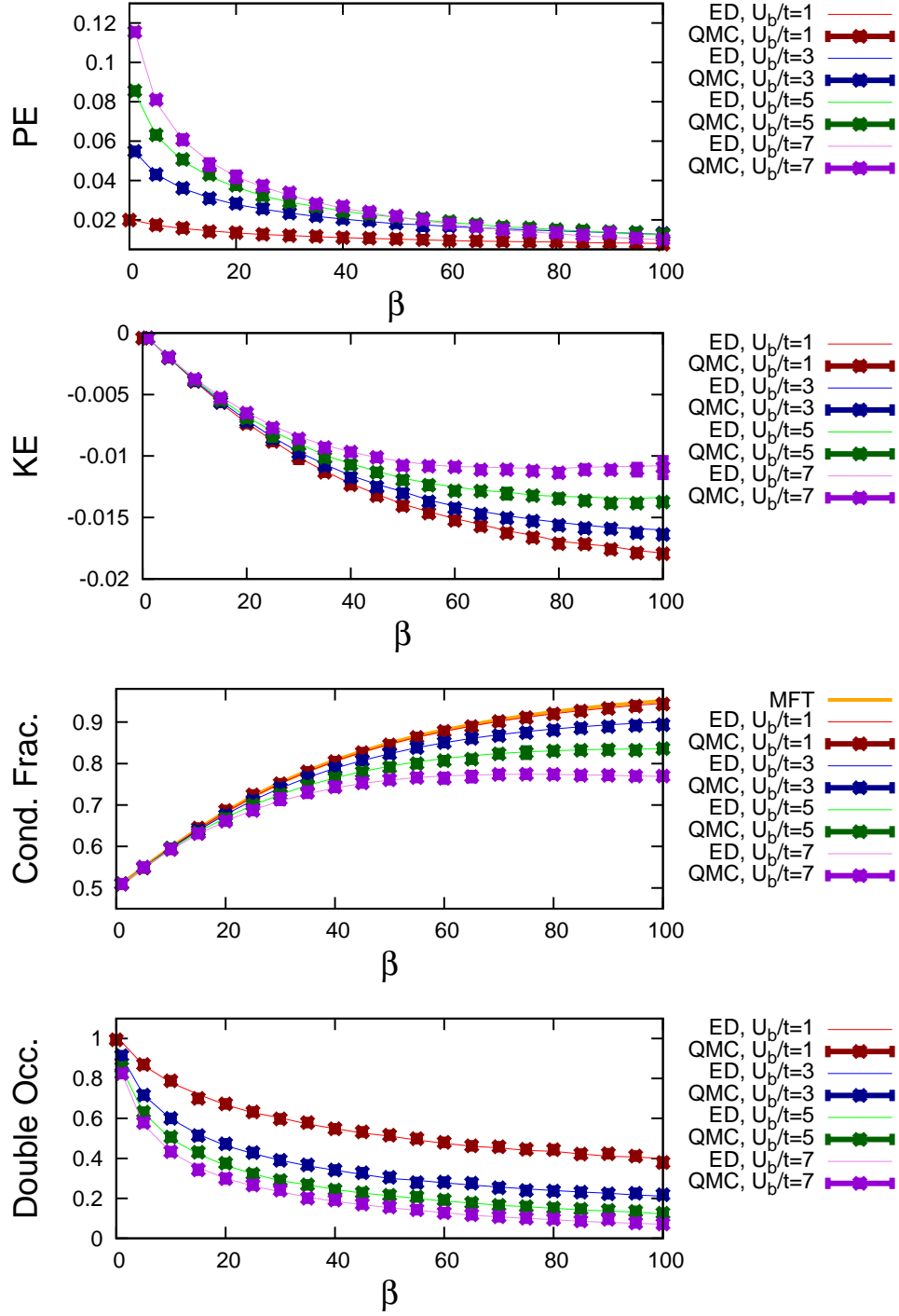}\end{center}
\caption{2-Site Bose-Fermi-Hubbard model kinetic energies (KE), potential energies (PE), condensate fractions, and double occupancies per site for varying $U_{b}=C$, $t_{b}=t_{f}=0.01$, and $\langle n_{b} \rangle=\langle n_{f} \rangle=1$ using both ED and BF-AFQMC. $\beta$ denotes the inverse temperature. BF-AFQMC results are in exact agreement with those from ED.}
\label{2SiteBFHAll}
\end{figure} 

\begin{figure}
\begin{center}\includegraphics[width=160mm]{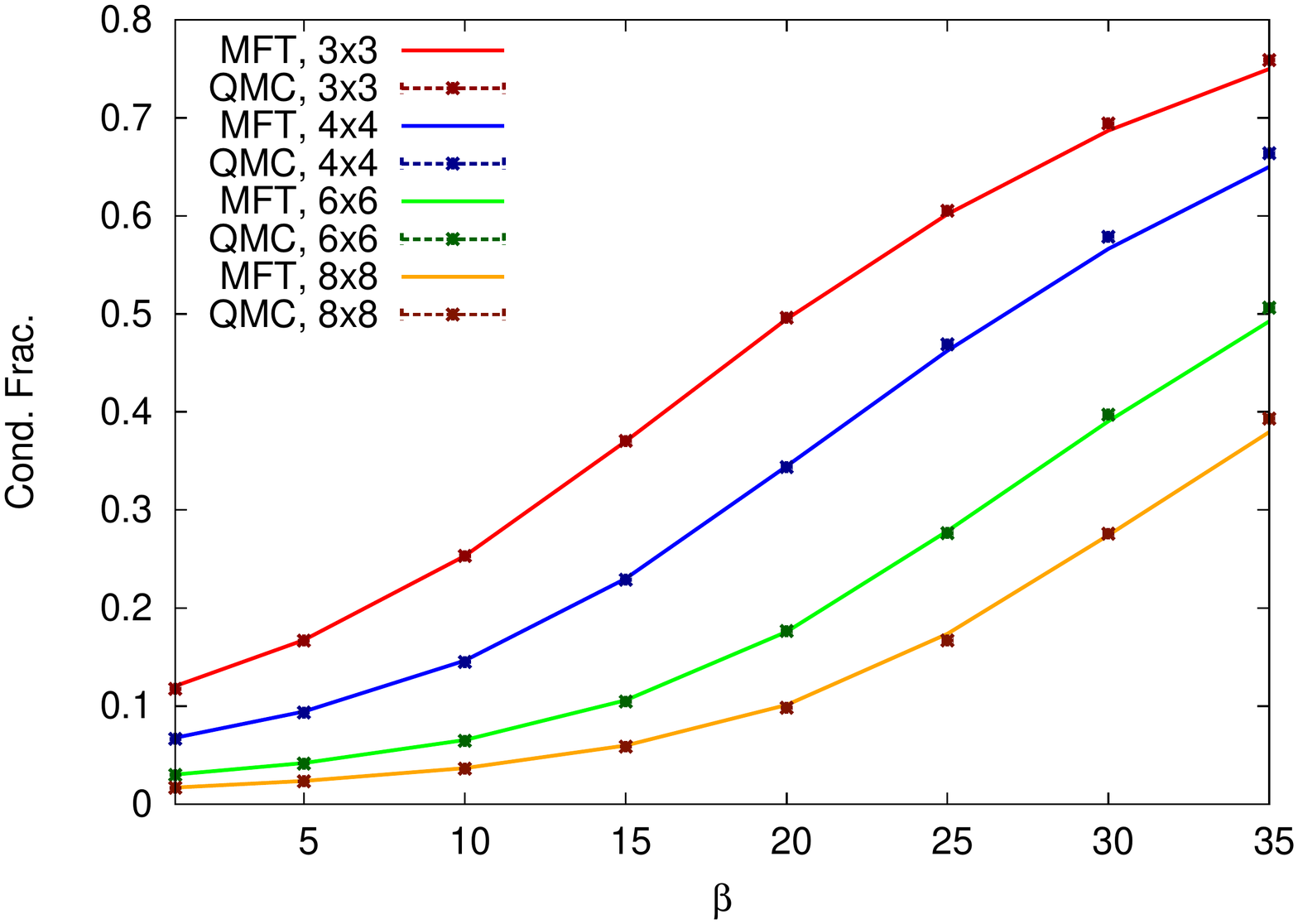}\end{center}
\caption{QMC and MFT condensate fractions for the 2D Bose-Fermi-Hubbard model at $U_{b}/t=C/t=0.5$, $t_{b}=t_{f}=0.01$, and $\langle n_{b} \rangle=\langle n_{f} \rangle=1$. $\beta$ denotes the inverse temperature. Good agreement is found between QMC and MFT at high temperatures.}
\label{2DBFMFTQMC}
\end{figure} 

\end{widetext}


\begin{thebibliography}{99}

\bibitem{ColdAtomBloch}
I. Bloch, Nature Physics. {\bf 1}, 23 (2005). 

\bibitem{ColdAtomLewenstein}
M. Lewenstein, A. Sanpera, V. Ahufinger, B. Damski, A. Sen(De), and U. Sen, Advances in Physics. {\bf 56}, 243 (2007). 
  
\bibitem{SFMI}
M. Greiner, O. Mandel, T. Esslinger, T.W. Hansch, and I. Bloch, Nature. {\bf 415}, 40 (2002). 
  
\bibitem{ColdAtomFermionsTruscott}
A.G. Truscott, K.E. Strecker, W.I. McAlexander, G.B. Partridge, and R.G. Hulet, Science. {\bf 291}, 2570 (2001). 

\bibitem{ColdAtomFermionsRom}
T. Rom, Th. Best, D. van Oosten, U. Schneider, S. Folling, B. Paredes, and I. Bloch, Nature. {\bf 444}, 733 (2006). 

\bibitem{Inouye}
S. Inouye, J. Goldwin, M.L. Olsen, C. Ticknor, J.L. Bohn, and D.S. Jin, Phys. Rev. Lett. {\bf 93}, 183201 (2004). 

\bibitem{Goldwin}
J. Goldwin, S. Inouye, M.L. Olsen, B. Newman, B.D. DePaola, and D.S. Jin, Phys. Rev. Lett. {\bf 70}, 021601 (2004). 

\bibitem{Ospelkaus1}
S. Ospelkaus, C. Ospelkaus, O. Wille, M. Succo, P. Ernst, K. Sengstock, and K. Bongs, Phys. Rev. Lett. {\bf 96}, 180403 (2006). 

\bibitem{Ospelkaus2}
C. Ospelkaus, S. Ospelkaus, L. Humbert, P. Ernst, K. Sengstock, and K. Bongs, Phys. Rev. Lett. {\bf 97}, 120402 (2006). 

\bibitem{Ospelkaus3} 
S. Ospelkaus, C. Ospelkaus, L. Humbert, K. Sengstock, and K. Bongs, Phys. Rev. Lett. {\bf 97}, 120403 (2006). 

\bibitem{Ospelkaus4}
C. Ospelkaus, S. Ospelkaus, K. Sengstock, and K. Bongs, Phys. Rev. Lett. {\bf 96}, 020401 (2006). 

\bibitem{Gunter}
K. Gunter, T. Stoferle, H. Moritz, M. Kohl, and T. Esslinger, Phys. Rev. Lett. {\bf 96}, 180402 (2006).  

\bibitem{Schreck}
F. Schreck, L. Khaykovich, K.L. Corwin, G. Ferrari, T. Bourdel, J. Cubizolles, and C. Salomon, Phys. Rev. Lett. {\bf 87}, 080403 (2001). 

\bibitem{Hadzibabic}
Z. Hadzibabic, C.A. Stan, K. Dieckmann, S. Gupta, M.W. Zweirlein, A. Gorlitz, and W. Ketterle, Phys. Rev. Lett. {\bf 88}, 160401 (2002). 

\bibitem{Heiselberg}
H. Heiselberg, C.J. Pethick, H. Smith, and L. Viverit, Phys. Rev. Lett. {\bf 85}, 2418 (2000). 

\bibitem{Efremov}
D.V. Efremov and L. Viverit, Phys. Rev. B. {\bf 65}, 134519 (2002). 

\bibitem{Illuminati}
F. Illuminati and A. Albus, Phys. Rev. Lett. {\bf 93}, 090406 (2004). 

\bibitem{BFPhaseDiagramLewenstein}
M. Lewenstein, L. Santos, M.A. Baranov, and H. Fehrmann, Phys. Rev. Lett. {\bf 92}, 050401 (2004). 
   
\bibitem{BFPhaseDiagramFehrmann}
H. Fehrmann, M.A. Baranov, B. Damski, M. Lewenstein, and L. Santos, Optics Comm. {\bf 243}, 23 (2004).

\bibitem{Zirbel}
J.J. Zirbel, K.-K. Ni, S. Ospelkaus, T.L. Nicholson, M.L. Olsen, P.S. Julienne, C.E. Wieman, J. Ye, and D.S. Jin, Phys. Rev. A. {\bf 78}, 013416 (2008). 

\bibitem{DeMille}
D. DeMille, Phys. Rev. Lett. {\bf 88}, 067901 (2008). 

\bibitem{OspelkausReview}
C. Ospelkaus and S. Ospelkaus, J. Phys. B. {\bf 41}, 203001 (2008). 

\bibitem{Buchler}
H.P. Buchler and G. Blatter, Phys. Rev. Lett. {\bf 91}, 130404 (2003). 

\bibitem{Kuklov}
A.B. Kuklov and B.V. Svistunov, Phys. Rev. Lett. {\bf 90}, 100401 (2003). 

\bibitem{Albus}
A.P. Albus, S.A. Gardiner, F. Illuminati, and M. Wilkens, Phys. Rev. A. {\bf 65}, 053607 (2002). 

\bibitem{Cramer}
M. Cramer, J. Eisert, and F. Illuminati, Phys. Rev. Lett. {\bf 93}, 190405 (2004). 

\bibitem{Mathey}
L. Mathey, D.-W. Wang, W. Hofstetter, M.D. Lukin, and E. Demler, Phys. Rev. Lett. {\bf 93}, 120404 (2004). 

\bibitem{Zujev}
A. Zujev, A. Baldwin, R.T. Scalettar, V.G. Rousseau, P.J.H. Denteneer, and M. Rigol, Phys. Rev. A. {\bf 78}, 033619 (2008). 

\bibitem{Pollet}
L. Pollet, C. Kollath, U. Schollwock, and M. Troyer, Phys. Rev. A. {\bf 77}, 023608 (2008). 

\bibitem{Varney} 
C.N. Varney, V.G. Rousseau, and R. T. Scalettar, Phys. Rev. A. {\bf 77}, 041608 (2008). 

\bibitem{Hebert1}
F. Hebert, F. Haudin, L. Pollet, and G. G. Batrouni, Phys. Rev. A. {\bf 76}, 043619 (2007). 

\bibitem{Hebert2}
F. Hebert, G. G. Batrouni, X. Roy, and V.G. Rousseau, Phys. Rev. B. {\bf 78}, 184505 (2008). 

\bibitem{Schollwock}
U. Schollwock, Rev. Mod. Phys. {\bf 77}, 259-315 (2005). 

\bibitem{Kotliar}
A. Georges, G. Kotliar, W. Krauth, and M.J. Rozenberg, Rev. Mod. Phys. {\bf 68}, 13 (1996). 

\bibitem{Byczuk}
K. Byczuk and D. Vollhardt, Annalen Der Physik. {\bf 18}, 622 (2009). 

\bibitem{Titvinidze1}
I. Titvinidze, M. Snoek, and W. Hofstetter, Phys. Rev. Lett. {\bf 100}, 100401 (2008). 

\bibitem{Titvinidze2}
I. Titvinidze, M. Snoek, and W. Hofstetter, Phys. Rev. B. {\bf 79}, 144506 (2009). 

\bibitem{AndersBFDMFT}
P. Anders, P. Werner, M. Troyer, M. Sigrist, and L. Pollet, arXiv:1203.6359. 

\bibitem{AndersBDMFT1}
P. Anders, E. Gull, L. Pollet, M. Troyer, and P. Werner, Phys. Rev. Lett. {\bf 105}, 096402 (2010). 

\bibitem{AndersBDMFT2} 
P. Anders, E. Gull, L. Pollet, M. Troyer, and P. Werner, New J. Phys. {\bf 13}, 075013 (2011). 

\bibitem{Prokofev1}
N.V. Prokofev, B.V. Svistunov, and I.S. Tupitsyn, Physics Letters A. {\bf 238}, 253 (1998). 

\bibitem{Prokofev2}
N.V. Prokofev, B.V. Svistunov, and I.S. Tupitsyn, JETP. {\bf 87}, 310 (1998). 

\bibitem{Batrouni1}
G.G. Batrouni and R.T. Scalettar, Phys. Rev. B. {\bf 46}, 9051 (1992). 

\bibitem{Batrouni2}
G.G. Batrouni and R.T. Scalettar, Comp. Phys. Comm. {\bf 97}, 63 (1996).  
   
\bibitem{VanHoucke}
K. Van Houcke, E. Kozik, N. Prokofev, and B. Svistunov, Physics Procedia. {\bf 6}, 95 (2010). 
   
\bibitem{Prokofev3}
N.V. Prokofev and B.V. Svistunov, Phys. Rev. B. {\bf 77}, 125101 (2008). 

\bibitem{Phaseless}
S. Zhang and H. Krakauer, Phys. Rev. Lett. {\bf 90}, 136401 (2003).

\bibitem{BSS1}
R. Blankenbecler, D.J. Scalapino, and R.L. Sugar, Phys. Rev. D. {\bf 24}, 2278 (1981). 
   
\bibitem{BSS2}
D.J. Scalapino and R.L. Sugar, Phys. Rev. Lett. {\bf 46}, 519 (1981). 
   
\bibitem{BSS3}
D.J. Scalapino and R.L. Sugar, Phys. Rev. B. {\bf 24}, 4295 (1981).

\bibitem{SignProblem}
E. Loh, J.E. Gubernatis, R.T. Scalettar, S.R. White, D.J. Scalapino, and R.L. Sugar, Phys. Rev. B. {\bf 41}, 9301 (1990). 

\bibitem{Sugiyama}
G. Sugiyama and S.E. Koonin, Ann. Phys. (N.Y.) {\bf 168}, 1 (1986). 

\bibitem{Bai}
Z. Bai, W. Chen, R. Scalettar, and I. Yamazaki. Numerical Methods for Quantum Monte Carlo Simulations of the Hubbard Model. World Scientific, (2007). 

\bibitem{SuzukiTrotter1}
M. Suzuki, Commun. Math. Phys. {\bf 51}, 183 (1976). 

\bibitem{SuzukiTrotter2}
H.F. Trotter, Proc. Am. Math. Soc. {\bf 10}, 545 (1959).  

\bibitem{HSTransform1}
G.M. Buendia, Phys. Rev. B. {\bf 33}, 3519 (1986). 
   
\bibitem{HSTransform2}
J.E. Hirsch, Phys. Rev. B. {\bf 28}, 4059 (1983).

\bibitem{CPMC1}
S. Zhang, Phys. Rev. Lett. {\bf 83}, 2777 (1999). 
   
\bibitem{CPMC2}
S. Zhang, J. Carlson, and J.E. Gubernatis, Phys. Rev. Lett. {\bf 74}, 3652 (1995). 

\bibitem{Chang2008}
Chia-Chen Chang and Shiwei Zhang, Phys. Rev. B. {\bf 78}, 165101 (2008).

\bibitem{alsaidi1}
W. A. Al-Saidi, Shiwei Zhang, and Henry Krakauer, J.~Chem.~Phys.~{\bf 124}, 224101 (2006).

\bibitem{alsaidi2}
W. A. Al-Saidi, Shiwei Zhang, and Henry Krakauer, J.~Chem.~Phys.~{\bf 127}, 144101 (2007). 

\bibitem{Purwanto1}
W. Purwanto and S. Zhang, Phys. Rev. E. {\bf 70}, 056702 (2004). 

\bibitem{Purwanto2}
W. Purwanto and S. Zhang, Phys. Rev. A. {\bf 72}, 053610 (2005). 

\bibitem{ZhangReview}
S. Zhang. Quantum Monte Carlo Methods for Strongly Correlated Electron Systems in \emph{Theoretical Methods for Strongly Correlated Electron Systems}. Springer-Verlag, (2003). 

\bibitem{ED}
A. Weibe and H. Fehske. Exact Diagonalization Techniques, Lecture Notes Physics. {\bf 739} 529-544 (2008). 

\bibitem{ALPS}
B. Bauer, et al., J. Stat. Mech. {\bf 2011}, 5001 (2011). 

\bibitem{Fetter}
A.L. Fetter and J. D. Walecka, Quantum Theory of Many Particle Systems. Dover Books: Mineola, NY (2003). 
   
\bibitem{Hirsch}
J.E. Hirsch, Phys. Rev. B. {\bf 31}, 4403 (1985).    

\bibitem{Stabilization}
S.R. White, D.J. Scalapino, R.L. Sugar, E. Loh, J.E. Gubernatis, and R.T. Scalettar, Phys. Rev. B. {\bf 40}, 506 (1989). 

\bibitem{DoubleOccupancy}
Y. Khorramzadeh, F. Lin, and V.W. Scarola, Phys. Rev. A. {\bf 85}, 043610 (2012). 

\bibitem{Klich}
I. Klich, arXiv:cond-mat/0209642v1 (2002). 
   
\bibitem{Balian}
R. Balian and E. Brezin, Nuovo Cimento. {\bf 64}, 37 (1969). 
   
\bibitem{Gubernatis}
J.E. Gubernatis and X.Y. Zhang, International Journal of Modern Physics C - Physics and Computers. {\bf 5}, 599 (1994).   

\bibitem{BoseHubbard}
M.P.A. Fisher, P.B. Weichman, G. Grinstein, and D.S. Fisher, Phys. Rev. B.~{\bf 40}, 546, (1989). 

\bibitem{HamannFahy}
Hamann, D. R. and Fahy, S. B., Phys. Rev. B.~{\bf 41}, 11352 (1990). 
\COMMENTED{
@article{PhysRevB.41.11352,
  title = {Energy measurement in auxiliary-field many-electron calculations},
  author = {Hamann, D. R. and Fahy, S. B.},
  journal = {Phys. Rev. B},
  volume = {41},
  issue = {16},
  pages = {11352--11363},
  year = {1990},
  month = {Jun},
  doi = {10.1103/PhysRevB.41.11352},
  url = {http://link.aps.org/doi/10.1103/PhysRevB.41.11352},
  publisher = {American Physical Society}
}
}

\end{thebibliography}
\end{document}